\documentclass[ twocolumn,aps,prd,   
               preprintnumbers,numbers,sort&compress,
               nofootinbib,
                            showpacs,
               colorlinks,
               linkcolor=blue,
               citecolor=blue]{revtex4-1}
   \newcommand{\exclude}[1]{}

\usepackage{graphicx,amsmath,amssymb,bm}
\usepackage{psfrag}
 \usepackage{feynmp}
\usepackage{hyperref}
\usepackage{graphicx}
\usepackage{subcaption}
\usepackage{xcolor} 

\newcommand{\beq}{\begin{equation}}
\newcommand{\eeq}{\end{equation}}
\newcommand{\be}{\begin{eqnarray}}
\newcommand{\ee}{\end{eqnarray}}

\def\ra{\rangle}
\def\la{\langle}

\def\rmd{\mathrm{d}}

\begin{document} 

\title{DAMA/LIBRA annual modulation and  Axion Quark Nugget   Dark Matter Model} 
\author{Ariel   Zhitnitsky}
\affiliation{ Department of Physics \& Astronomy, University of British Columbia, 
Vancouver, B.C. V6T 1Z1, Canada} 

\begin{abstract}
The DAMA/LIBRA experiment shows $9.5 \sigma$ evidence for an annual modulation  in the $(1-6)~ {\rm keV}$ energy range, strongly suggesting that the   observed modulation has the dark matter origin. However, the conventional interpretation in terms of WIMP-nucleon  interaction   is excluded by other experiments.    We propose an alternative source of modulation in the form of neutrons, which have been liberated   from surrounding material.
 Our computations are based on the so-called Axion Quark Nugget (AQN) dark matter model, which was originally invented long ago  to explain the similarity between the dark and visible cosmological matter densities, i.e.  $ \Omega_{\rm dark} \sim \Omega_{\rm visible}$. 
 \exclude{ In the AQN framework the nuggets could be 
made of matter as well as antimatter.  Annihilation of antimatter nuggets  with visible matter in the Earth's  interior    inevitably produces the neutrinos  when the nuggets  pass through  the Earth.  The corresponding neutrino flux is obviously a subject of  the annual variation, which eventually generates the DAMA/LIBRA annual modulation in the AQN framework. } In our proposal  the annual modulation is shown to be   generated in keV energy range which is consistent with DL observation in  $(1-6)~ {\rm keV}$ range. This keV energy scale  in our proposal   is  mostly  determined by spectral properties of the neutrinos emitted by the AQN dark matter particles, while the absence of the modulation with  energies above 6 keV 
is explained   by a  sharp cutoff in  the neutrino's energy spectrum  at $\sim 15$ MeV.  This proposal can be directly tested by    COSINE-100, ANAIS-112,   CYGNO  and other  similar experiments.  It can be also tested by studying  the correlations between the signals from these  experiments and the signatures  from drastically different detectors   designed for  studies  of the   infrasonic  or seismic events using  such instruments as Distributed Acoustic Sensing (DAS). 
\end{abstract}
\maketitle

\section{Introduction}
\label{introduction}
The DAMA/LIBRA (DL) experiment \cite{Bernabei:2013xsa,Bernabei:2014tqa,Bernabei:2016bkl,Bernabei:2018yyw}
claims the observation for an annual modulation  in the $(1-6)~ {\rm keV}$ energy range  at $9.5 \sigma$ C.L. 
 The C.L. even higher ($12.9 \sigma$) for $(2-6)~ {\rm keV}$ energy range  when DAMA/NaI and DL-phase1
 can be combined with DL-phase2 results.  The measured period   ($0.999\pm 0.001$) year and phase ($145\pm 5$) 
   strongly indicate to the 
 dark matter (DM) origin of the modulation, the phenomenon which was    originally suggested in \cite{Freese:1987wu}, see also   review paper \cite{Freese:2012xd}.
 
 However the annual modulation observed by DL is excluded by other direct detection experiments if interpreted in terms of the
 WIMP-nuclei interactions. 
 This  motivated a number of alternative explanations for the DL signal. 
 In the present work we argue  that the modulation observed by DL is due to the {\it neutrons} surrounding the detector. In this respect 
 our proposal is   similar to the previous suggestions \cite{Ralston:2010bd,Nygren:2011xu,Blum:2011jf,Davis:2014cja}
 where the authors agued that the induced neutrons (which have  been  liberated from material surrounding the detector)  
 may be responsible for the observed annual modulation. 
 
 Our proposal is drastically different from previous  suggestions \cite{Ralston:2010bd,Nygren:2011xu,Blum:2011jf,Davis:2014cja} in one but crucial aspect: the neutrons in our case are induced by neutrinos emitted by the  Axion Quark Nugget (AQN) dark matter particles.  Therefore, the annual modulation observed by DL has  truly genuine DM origin, though it is manifested indirectly in our framework through the following chain:
 \be
 \label{links}
 {\rm AQN}\rightarrow{\rm (neutrinos)}\rightarrow{\rm (surrounding~neutrons)}\rightarrow {\rm DL}.~~~~
 \ee 
 In this framework the modulation should  obviously show  a proper period of 1 year with proper phase as the neutrinos from (\ref{links}) are originated from dark matter nuggets, and the corresponding time variation will be obviously  transferred to the modification of the neutron flux,  which eventually generates the modulation observed by DL. 
 
  One should emphasize that the emission features of the neutrinos emitted by AQNs such as the intensity and spectrum
 (which ultimately determines   the $(1-6)~ {\rm keV}$ energy recoil  for  the observed DL annual modulation) have been computed in the AQN model long ago for  completely different purposes, and we will   use exactly  the same original parameters of the model without any intension to modify them to fit  the DL observations.  
 
 We overview the basic ideas of  the AQN model in the next section \ref{sec:QNDM}. 
One should emphasize that this model is consistent with all available cosmological, astrophysical, satellite and ground based constraints, where AQNs could leave a detectable electromagnetic signature. While the model was initially invented to explain the observed relation $\Omega_{\rm dark}\sim \Omega_{\rm visible}$, it may also explain a number of other (naively unrelated) phenomena, such as the  excess of galactic emission in different frequency bands.  
 The AQN model may also resolve  other, naively unrelated  astrophysical mysteries.
 It includes, but not limited:  the so-called ``Primordial Lithium Puzzle" \cite{Flambaum:2018ohm},  the so-called ``The Solar Corona Mystery"  \cite{Zhitnitsky:2017rop,Raza:2018gpb}, the recent EDGES observations   \cite{Lawson:2018qkc}, and unexpected 
annual modulation in    x-rays in $(2-6)~ {\rm keV}$ energy band  observed by XMM -Newton observatory \cite{Ge:2020cho},  among many others. 
These cosmological puzzles could  be resolved within AQN framework with the  same set of physical parameters to be used  in the present work to explain the annual modulation observed by DL, without fitting or modifications of any of them.

Our main results are formulated in  section \ref{AQN-neutrino} where we estimate the  intensity of modulation  due to the neutrino flux emitted by dark matter nuggets  when they traverse  through the earth. This section is separated to  4 different subsections, \ref{1}, \ref{2}, \ref{3} and \ref{4} according to 4 different elements of the proposal (\ref{links}). We use precisely the same set of parameters obtained previously for a different purpose, as reviewed in the Appendix \ref{cfl}.   We further estimate the neutrino- induced neutron's flux in surrounding rocks which according to the proposal (\ref{links}) is the source of the observed DL modulation signal. 

In section \ref{comments} we  comment on DL arguments \cite{Bernabei:2013xsa,Bernabei:2014tqa,Bernabei:2016bkl,Bernabei:2018yyw} suggesting  the  irrelevance of the induced neutron flux  (due to muons and  neutrinos). We explicitly show why DL arguments are not applicable for  our  proposal (\ref{links}) with its  truly genuine DM nature, though manifested  indirectly.  

In subsection     \ref{previous}
we   comment on  a number of  experiments which exclude  the DL results, while in subsection \ref{recent}  we make few  comments on recent results by  COSINE-100    \cite{Adhikari:2018ljm,Adhikari:2018fmv,Adhikari:2019off} and ANAIS-112 \cite{Amare:2019jul}, and recent proposal CYGNO 
  \cite{CYGNO:2019aqp}   which have been largely  motivated by   the DL observations. Finally, in subsections \ref{future} and \ref{correlations} we suggest few  novel  tests which could unambiguously support or rule out the proposal (\ref{links}).
 
  \section{Axion Quark nugget   DM model}
\label{sec:QNDM}
In this section we overview the AQN model. Specifically, in subsection \ref{basics} we list most important and   very generic features of the framework, which do not depend on specific  parameters of the model.
In   subsection \ref{sec:Constraints} we  overview the observational constraints on a single fundamental  parameter of this framework, the average baryon charge of the nugget  $\la B\ra$ which enters all the observables. Finally, in subsection \ref{sec:extra} we highlight a number of other model-dependent properties, such as ionization features of the AQNs, their survival pattern,  their annihilation rate and some other questions which are relevant for the present studies.

\subsection{Generic  features of the AQN model}\label{basics}
The idea that the dark matter may take the form of composite objects of 
standard model quarks in a novel phase goes back to quark nuggets  \cite{Witten:1984rs}, strangelets \cite{Farhi:1984qu}, nuclearities \cite{DeRujula:1984axn},  see also review \cite{Madsen:1998uh} with large number of references on the original results. 
In the models \cite{Witten:1984rs,Farhi:1984qu,DeRujula:1984axn,Madsen:1998uh}  the presence of strange quark stabilizes the quark matter at sufficiently 
high densities allowing strangelets being formed in the early universe to remain stable 
over cosmological timescales. 

The axion quark nuggets (AQN)  model   advocated in \cite{Zhitnitsky:2002qa} is conceptually similar, with the 
nuggets being composed of a high density colour superconducting (CS) phase.   
As with other high mass dark matter candidates  
  \cite{Witten:1984rs,Farhi:1984qu,DeRujula:1984axn,Madsen:1998uh} these objects are ``cosmologically dark" not through the weakness of their 
interactions but due to their small cross-section to mass ratio. 
As a result, the corresponding  constraints on this type of dark matter place a lower bound on their mass, rather than coupling constant.  

There are several additional elements in AQN model in comparison with the older well-known and well-studied constructions \cite{Witten:1984rs,Farhi:1984qu,DeRujula:1984axn,Madsen:1998uh}:
\begin{itemize} 
\item
First, there is an additional stabilization factor  provided by the axion domain walls (with a QCD substructure)
which are copiously produced during the QCD transition and which help to alleviate a number of 
the problems inherent in the older models\footnote{\label{first-order}In particular, in the original proposal the first order phase transition was the required feature of the construction.  However it is known  that the QCD transition is a crossover rather than the first order phase transition. It should be contrasted with  AQN framework
when  the first order phase transition is not required  as the axion domain wall plays the role of the squeezer.  Furthermore, it had been argued that the nuggets   \cite{Witten:1984rs,Farhi:1984qu,DeRujula:1984axn,Madsen:1998uh} 
are likely to evaporate on the Hubble time-scale even if they were formed. In the AQN framework 
the  fast evaporation arguments    do not apply    because the  vacuum ground state energies in the  CS  and    hadronic phases   are drastically different.}.  
\item The core of the AQN is in CS phase, which implies that the two systems (CS and hadronic) can coexist only in the presence of the  external pressure which is provided by the axion domain wall. It should be contrasted with original models  \cite{Witten:1984rs,Farhi:1984qu,DeRujula:1984axn,Madsen:1998uh} which  must  be stable at zero external pressure.
\item Another  
crucial    additional  element in the proposal      is that the nuggets could be 
made of matter as well as  antimatter in this framework. 
\end{itemize} 
The direct consequence of this feature is that the dark matter  density  
   $\Omega_{\rm dark}$ and the baryonic matter density $ \Omega_{\rm visible}$ will automatically assume the same order of magnitude    $\Omega_{\rm dark}\sim \Omega_{\rm visible}$ without any fine tunings, and irrespectively to  any specific details of the model, such as the axion mass $m_a$ or size of the nuggets $R$. Precisely this fundamental consequence of the model was the main  motivation for its construction.

The presence of a large amount of antimatter in the form of high density AQNs leads to  many  
 observable consequences   as a result of  possible (but, in general, very rare) annihilation events between antiquarks from AQNs and  baryons from visible Universe. We   highlight  below 
  the  basic  results,  accomplishments and constraints of this model.

Let us recapitulate the original motivation for this model: 
it is commonly  assumed that the Universe 
began in a symmetric state with zero global baryonic charge 
and later (through some baryon number violating process, non- equilibrium dynamics, and $\cal{CP}$ violation effects, realizing three  famous   Sakharov's criteria) 
evolved into a state with a net positive baryon number.

As an 
alternative to this scenario we advocate a model in which 
``baryogenesis'' is actually a charge separation (rather than charge generation) process 
in which the global baryon number of the universe remains 
zero at all times.  In this model the unobserved antibaryons come to comprise 
the dark matter in the form of dense nuggets of   antiquarks and gluons in CS phase.  
The result of this ``charge separation"  process is two populations of AQN carrying positive and 
negative baryon number. In other words,  the AQN may be formed of either matter or antimatter. 
However, due to the global  $\cal CP$ violating processes associated with $\theta_0\neq 0$ during 
the early formation stage,  the number of nuggets and antinuggets 
  will be different\footnote{This source of strong ${\cal CP}$ violation is no longer 
available at the present epoch as a result of the dynamics of the axion, 
 which  remains the most compelling resolution of the strong ${\cal CP}$ problem, see original papers 
 on PQ symmetry \cite{1977PhRvD..16.1791P}, Weinberg-Wilczek axion \cite{1978PhRvL..40..223W,1978PhRvL..40..279W},
 KSVZ invisible  axion \cite{KSVZ1,KSVZ2} and DFSZ invisible axion  \cite{DFSZ1,DFSZ2} models. See also  recent reviews \cite{vanBibber:2006rb, Asztalos:2006kz,Sikivie:2008,Raffelt:2006cw,Sikivie:2009fv,Rosenberg:2015kxa,Marsh:2015xka,Graham:2015ouw,Ringwald:2016yge,Irastorza:2018dyq}.}.
 This difference is always an order of one effect   irrespectively to the 
parameters of the theory, the axion mass $m_a$ or the initial misalignment angle $\theta_0$.
We refer to the original papers   \cite{Liang:2016tqc,Ge:2017ttc,Ge:2017idw,Ge:2019voa} devoted to the specific questions  related to the nugget's formation, generation of the baryon asymmetry, and 
survival   pattern of the nuggets during the evolution in  early Universe with its unfriendly environment. The only comment we would like to make here is that 
         the disparity  between nuggets $\Omega_{N}$ and antinuggets $\Omega_{\bar{N}}$ generated due to  $\cal CP$ violation  unambiguously implies that  the  baryon contribution $\Omega_{B}$ must be the same order of magnitude as $\Omega_{\bar{N}}$  and $\Omega_{N}$ because all these contributions are proportional to one and the same fundamental dimensional parameter $\Lambda_{\rm QCD}$. If these processes are not fundamentally related the two components $ \Omega_{\rm dark}$ and $\Omega_{\rm visible}$ could easily assume drastically  different scales.
 This represents a  precise mechanism of how the  ``baryogenesis" can be  replaced by  the ``charge separation"  processes  in the AQN framework.

The remaining antibaryons in the early universe plasma then 
annihilate away leaving only the baryons whose antimatter 
counterparts are bound in the excess of antiquark nuggets and are thus 
unavailable for fast annihilation. All asymmetry effects   are order of
one which eventually  results  in similarities for all 
  components, visible and dark, i.e.
\be
\label{Omega}
 \Omega_{\rm dark} \sim \Omega_{\rm visible}, ~~~  \Omega_{\rm dark}\approx [\Omega_{N}+ \Omega_{\bar{N}}],
\ee
as they are both proportional to the same fundamental $\Lambda_{\rm QCD} $ scale,  
and they both are originated at the same  QCD epoch. 
In particular, the observed 
matter to dark matter ratio $\Omega_{\rm dark} \simeq 5\cdot \Omega_{\rm visible}$ 
corresponds to a scenario in which the number of antinuggets is larger than the 
number of nuggets by a factor of $ (\Omega_{\bar{N}}/ \Omega_{N}) \sim$ 3/2 at the end of nugget formation. 
It is important to emphasize that the AQN mechanism is not sensitive to the axion mass $m_a$ and it is capable to saturate (\ref{Omega}) itself without any other additional contributors. It should be contrasted with conventional axion production mechanisms when the corresponding  contribution  scales as $\Omega_{\rm axion}\sim m_a^{-7/6}$. This scaling  unambiguously implies that the axion mass must be fine-tuned  $m_a\simeq 10^{-5} {\rm eV}$
 to  saturate the DM density today  while larger axion mass will contribute very little to $\Omega_{\rm dark}$. The relative role  between the direct axion contribution $\Omega_{\rm axion}$ and the AQN contribution to $\Omega_{\rm dark}$ as a function of mass $m_a$ has been evaluated in \cite{Ge:2017idw}, see Fig. 5 in that paper. 
   
 Unlike conventional dark matter candidates, such as WIMPs 
(Weakly interacting Massive Particles) the dark-matter/antimatter
nuggets are strongly interacting and macroscopically large.  
However, they do not contradict any of the many known observational
constraints on dark matter or
antimatter  due to the following  main reasons~\cite{Zhitnitsky:2006vt}:
\begin{itemize} 
\item They are absolutely stable configurations on cosmological scale as the pressure due to the axion domain  wall (with the QCD substructure)   is equilibrated by the Fermi pressure. Furthermore, it has been shown that the AQNs survive an unfriendly environment of early Universe, before and after BBN epoch \cite{Ge:2019voa}. The majority of the AQNs also survive such violent events as  the galaxy formation and star formation;
\item They carry a huge (anti)baryon charge 
$|B|  \gtrsim 10^{25}$, and so have an extremely tiny number
density; 
\item The nuggets have nuclear densities, so their effective cross section $\sigma\sim R^2$
is small relatively to its mass,  $\sigma/M \sim 10^{-10}$ ~cm$^2$/g. This  key ratio   is well below the typical astrophysical
and cosmological limits which are on the order of 
$\sigma/M<1$~cm$^2$/g ;
\item They have a large binding energy such that the baryon charge locked in the
nuggets is not available to participate in big bang nucleosynthesis
(BBN) at $T \sim 0.1$~MeV, and the basic BBN picture holds, with possible small corrections 
of order $\sim 10^{-10}$ which may in fact resolve the primordial lithium puzzle \cite{Flambaum:2018ohm};
\item The nuggets completely decouple from photons as a result of small $ \sigma/M \sim B^{-1/3}\ll 1$ ratio, such that conventional picture for structure formation holds. 
\item The nuggets do not modify conventional CMB analysis, with possible  small corrections  which, in fact, may resolve a tension  \cite{Lawson:2018qkc} between standard prediction and EDGES observation (stronger than anticipated 21 cm absorption features).
\end{itemize} 
 To reiterate: the weakness of the visible-dark matter interaction is achieved 
in this model due to the small geometrical factor $ {\sigma}/{M} \sim B^{-1/3}$ 
rather than due to a weak coupling of a new fundamental field to standard model particles. 
In other words, this small effective interaction $\sim \sigma/M \sim B^{-1/3}$ 
replaces a conventional requirement
of sufficiently weak interactions of the visible matter with WIMPs. 

\subsection{Mass distribution constraints}\label{sec:Constraints}
 One should emphasize that the AQN construction does not specify the size of the nuggets. In general, for   a given axion mass $m_a$ there is always a range of $B$ when a stable solution exists \cite{Zhitnitsky:2002qa}.  The average size of the nugget within this stability range scales as $R\propto m_a^{-1}$ while the baryon charge of the AQN  itself scales as $\la B\ra \propto R^3 \propto m_a^{-3}$.  
 In the AQN framework we treat $\langle B\rangle$ as a fundamental parameter 
to be constrained observationally. It is clear that    larger $\langle B\rangle$ values 
produce  weaker observational signals as $ {\sigma}/{M} \sim B^{-1/3}$. 
Furthermore, any such consequences assume the  largest values where the densities 
of both visible and dark matter are sufficiently high  such as in the 
core of the galaxy, the early universe, or the stars and planets. In other words, the nuggets behave like  
conventional cold dark matter in environments where the density of the visible matter is small, 
while they become interacting and radiating objects (i.e. effectively become visible matter)  
if they enter an environment of sufficiently large density.

As mentioned above the flux of AQN on the earth's surface is scaled by a factor of $B^{-1}$ 
and is thus suppressed for large nuggets,  see  eq.  
  (\ref{eq:D Nflux 3}) below. 
For this reason the experiments most relevant to AQN detection 
are not the conventional high sensitivity dark matter searches but detectors with the largest 
possible search area. For example it has been proposed \cite{Lawson:2010uz} that large scale cosmic ray detectors 
such as the Auger observatory of Telescope Array may be sensitive to the flux of AQN in an 
interesting mass range.  An obvious challenging problem with such studies is that the conventional cosmic ray detectors are designed to analyze  the time delays  measured in $\mu {\rm s}$ as cosmic rays are assumed to be moving  with the speed of light, while AQNs move with non-relativistic velocity $  v_{\rm AQN}  \simeq10^{-3}$ c. It obviously requires the latency time    to be measured   in ${\rm ms}$ range in order to study the correlated signals from AQN. The modern cosmic ray detectors are not designed  to analyze such long time correlations, and normally such signals would be rejected as a background. 

The strongest direct detection limit 
is  set by the IceCube Observatory's non-detection of a non-relativistic magnetic monopole
\cite{Aartsen:2014awd}. While the magnetic monopoles and the AQNs interact with material of the detector
in very different way, in both cases the interaction leads to the electromagnetic and hadronic cascades along the trajectory 
of AQN (or magnetic monopole) which must be observed by the  detector if such event occurs.  A non-observation of any such cascades puts the following  limit on the flux of heavy non-relativistic particles passing through the detector,  
\be
\label{direct}
\la B \ra > 3\cdot 10^{24} ~~~[{\rm direct ~ (non)detection ~constraint]}, 
\ee
where we assume   100 \% efficiency of the observation of the AQNs passing through IceCube Observatory, see Appendix A in ref.\cite{Lawson:2019cvy}.

 Similar limits are also 
derivable from the Antarctic Impulsive Transient Antenna  (ANITA) \cite{Gorham:2012hy} though 
this result   depends  on the details of radio band emissivity of the AQN. In the same work the author also  derives the   constraint arising from  potential contribution of the AQN annihilation events to earth's energy budget which 
requires $|B| > 2.6\times 10^{24}$ \cite{Gorham:2012hy}, which is also consistent with (\ref{direct}). 
 There is also a 
constraint on the flux of heavy dark matter with mass $M<55$g based on the non-detection of 
etching tracks in ancient mica \cite{Jacobs:2014yca}. It slightly touches the lower bound of the allowed window (\ref{direct}), but does not strongly constraint entire window  (\ref{window}) because the dominant portion of the AQNs lies well above  its lower limit   (\ref{direct}) assuming the mass distribution (\ref{distribution}) as discussed below. 

   The authors of ref.  \cite{Herrin:2005kb} use the Apollo data to constraint   the abundance of  quark nuggets  in the region of 10 kg to one ton. It has been argued that the contribution of such very heavy nuggets  must be at least an order of magnitude less than would saturate the dark matter in the solar neighbourhood \cite{Herrin:2005kb}. Assuming that the AQNs saturate the dark matter the constraint  \cite{Herrin:2005kb} can be reinterpreted   that at least 90\% of the AQNs must have the masses below 10 kg. 
   This constraint can be approximately expressed  in terms of the baryon charge as follows:
   \be
\label{apollo}
\la B \ra \lesssim   10^{28} ~~~[{\rm  Apollo~ constraint ]}. 
\ee
Therefore, indirect observational constraints (\ref{direct}) and (\ref{apollo}) suggest that if the AQNs exist and saturate the dark matter density today, the dominant portion of them   must reside in the window: 
\be
\label{window}
3\cdot 10^{24}\lesssim\la B \ra \lesssim   10^{28}~ [{\rm constraints~ from~ observations}].  ~~~
\ee 
 Completely different and independent observations  also suggest  that the galactic spectrum 
contains several excesses of diffuse emission the origin of which is not well established,  and  remains to be debated.
 The best 
known example is  the strong galactic 511~keV line. If the nuggets have a baryon 
number in the $\langle B\rangle \sim 10^{25}$ range they could offer a 
potential explanation for several of 
these diffuse components. It is very nontrivial consistency check that the required $\langle B\rangle$ to explain these excesses of the galactic diffuse emission  belongs to the same mass range as reviewed above. 
 For further details see the original works \cite{Oaknin:2004mn, Zhitnitsky:2006tu,Forbes:2006ba, Lawson:2007kp,Forbes:2008uf,Forbes:2009wg} with explicit  computations 
 of the galactic radiation  excesses  for varies frequencies, including excesses of the diffuse  x- and   $\gamma$- rays.  
In all these cases photon emission originates 
from the outer layer of the nuggets known as the electrosphere, and all intensities in different frequency bands are expressed in terms of a single parameter $\langle B\rangle$ such that all relatives intensities   are unambiguously fixed because  they are determined by the Standard Model (SM) physics.

Yet another AQN-related effect might be intimately linked to the so-called ``solar corona heating mystery".
The  renowned  (since 1939)  puzzle  is  that the corona has a temperature  
$T\simeq 10^6$K which is 100 times hotter than the surface temperature of the Sun, and 
conventional astrophysical sources fail to explain the extreme UV (EUV) and soft x ray radiation 
from the corona 2000 km  above the photosphere. Our comment here is that this puzzle  might  find its  
natural resolution within the AQN framework as recently argued in 
 \cite{Zhitnitsky:2017rop,Zhitnitsky:2018mav,Raza:2018gpb}.

In this scenario the AQN composed 
of antiquarks fully annihilate within the so-called transition region (TR) providing a total annihilation energy  
 which is very  close to the 
observed   EUV luminosity of $10^{27}$erg/s. The EUV emission is assumed  to be powered 
by impulsive heating events known as nanoflares conjectured by Parker long ago. The physical origin of the nanoflares  remains to be unknown. If our identification of these nanoflares 
with the  AQN annihilating events  is correct,  we may extract   the baryon charge  distribution $dN/dB$ for the AQNs because the energy distribution $dN/dE$ of the nanoflares has been previously modelled by the solar physics people to fit the observations, i.e.  
\be
  \label{distribution}
   {dN} \propto  E^{-\alpha} dE \propto B^{-\alpha}dB, 
  \ee
where  slop parameter $\alpha$ slightly varies for  different  solar models\footnote{One should comment here that the algebraic behaviour  for the   distribution   (\ref{distribution}) is a very generic feature of the percolation theory within AQN framework as recently argued in \cite{Ge:2019voa}.   However, a numerical  estimate of parameter $\alpha$ from the theoretical side requires a deep  understanding of the QCD phase transition dynamics at $\theta\neq 0$, when the axion domain walls are formed. This knowledge is not likely  to be available any time soon as the QCD lattice simulations cannot study a system with $\theta\neq 0$, which represents a well-known sign problem in the lattice community.}. 
It is a highly nontrivial consistency check that the typical nanoflare energy range $E_{\rm nano}\simeq (10^{22}-6\cdot 10^{25})~ {\rm erg}$
corresponds (within AQN framework) to the baryon charge window $3\cdot 10^{24}\lesssim B\lesssim  2\cdot 10^{28}$ which strongly overlaps with all presently available constraints (\ref{window}) on the AQN sizes as reviewed above. Precisely this highly nontrivial consistency check on size distribution along with (essentially model-independent) 
computation of the total luminosity of the EUV radiation $ \sim 10^{27} {\rm erg/s}$  from the solar corona, which is  consistent with observations,  gives us a strong confidence for the plausibility  of  the identification (\ref{distribution}) between the nanoflares conjectured by Parker long ago and the AQN annihilation events.
 
Encouraged by these consistency checks we adopted the AQN size distribution (\ref{distribution}) with parameter $\alpha$ being extracted from the heating corona studies for  all our subsequent Monte Carlo simulations including the computations of the AQN flux on Earth as given by   (\ref{eq:D Nflux 3}) to be discussed in great details in next section.

  \subsection{Few additional comments}\label{sec:extra} 
  Here we make few additional comments on  basic features of the AQNs   which are important for understanding of this  framework in general and for  specific estimates  (\ref{links}) relevant for DAMA/LIBRA signal in particular. 
  
  We start this overview with mentioning of the electric charge which AQNs may carry while propagating  in a media.
  It is normally assumed that all types of nuggets including the old version models  \cite{Witten:1984rs,Farhi:1984qu,DeRujula:1984axn,Madsen:1998uh}  are neutral at  zero temperature $T=0$ because the electrosphere made of leptons will be always formed even when the quark nuggets themselves are electrically charged (for example, due to the differences in quark's  masses).  However,  the neutrality will be lost due to the ionization at $T\neq 0$,  in which case  the  nuggets will esquire the positive charge due to the ionized electrons while the anti-matter nuggets will esquire the negative electric charge due to the ionized positrons. The corresponding charge $Q$ for AQNs can be estimated as  follows \cite{Zhitnitsky:2017rop,Flambaum:2018ohm}:
  \be
  \label{Q}
Q\simeq 4\pi R^2 \int^{\infty}_{z_0}  n(z)dz\sim \frac{4\pi R^2}{2\pi\alpha}\cdot \left(T\sqrt{2 m_e T}\right), 
  \ee
 where $n(z)$ is the density of the positrons in electrosphere which has been computed in the mean field approximation. In this estimate it is  assumed that the positrons with $p^2/(2m_e)< T$ will be stripped off the electrosphere as a result of high temperature $T$. These  loosely bound positrons 
 are localized mostly at outer regions of electrosphere at distances $z> z_0= (2 m_e T)^{-1/2}$ which motivates our cutoff in estimate (\ref{Q}). Numerically $Q\sim 10^8$ represents  very tiny portion $Q/B$   in comparison with the baryon charge $B\sim 10^{25}$ even for relatively high temperature $T\simeq 100$ eV in  the solar corona. These objects behave in all respects as neutral objects (for example   the cosmic magnetic field does not affect   the AQN's trajectory) because $Q/M \ll e/m_p$.   Nevertheless, a non-vanishing charge $Q$ may play a very important  role in some circumstances, such as propagation of AQNs in highly ionized plasma in solar corona   
 \cite{Zhitnitsky:2017rop,Zhitnitsky:2018mav,Raza:2018gpb}.  The non-vanishing charge Q may also  suppress  the primordial lithium abundance at $T\simeq 20 $ keV 
 due to the strong attraction between the negatively charged AQNs and positively charged  lithium nuclei \cite{Flambaum:2018ohm}.
  
  Our next comment is related to the survival pattern of the  AQNs. The comprehensive  studies on this matter can be found in the original work \cite{Ge:2019voa}. The only comment we would like to make here is that the dominant portion of the AQNs will survive the evolution of the Universe. However, very small portion  of the AQNs which is gravitationally  captured by stars and planets may drastically decrease their baryon charges and may even experience of   complete   annihilation\footnote{\label{matter-AQN}The nuggets made of matter can be  be stopped in dense environment such as neutron stars,  on the scales of order $L^{\rm stop}$ when the number of hits  is of order of  $B$, i.e. $\pi R^2 L^{\rm stop} \la \rho\ra \sim Bm_p$. However, in contrast with conventional nugget's models  \cite{Witten:1984rs,Farhi:1984qu,DeRujula:1984axn,Madsen:1998uh} the AQNs will not turn an entire star into a new phase because the CS phase in AQN is supported  by external pressure due to the axion domain wall, while the original nuggets   \cite{Witten:1984rs,Farhi:1984qu,DeRujula:1984axn,Madsen:1998uh}  are assumed to be stable objects at zero external pressure. The phenomenon of collecting the matter nuggets in the cores of stars/planets might  of interest by itself,  but it is not the topic of the present work devoted to antimatter AQNs   capable to produce neutrinos as a result of annihilation processes.}. The typical size $L$ of the region of the medium with density $\rho$ where the complete  annihilation occurs is estimated as $\sigma  \rho L\simeq B m_p$, where $\sigma$ is an effective cross section which could be much larger than  naive $\pi R^2$    due to the long-range  Coulomb interaction when $Q\neq 0$ according to (\ref{Q}). 
  This effect plays a very important  role in the solar corona  \cite{Zhitnitsky:2017rop,Zhitnitsky:2018mav,Raza:2018gpb}. All nuggets which are gravitationally  captured by the Sun will be completely annihilated.
  
  One may wonder what happens to the axions from AQNs which are now liberated and become propagating axions with average energy $\la E\ra\simeq 1.3 m_a$.  Small portion of these axions will be converted to photons in the background  of the solar magnetic field, and in principle can be observed on Earth. The effect is very small  though  even if one takes into account  the resonance condition  due to the   plasma effects   in the solar corona \cite{Liang:2018ecs}.

  In case of  the AQN traversing the Earth's interior only some small portion of the baryon charge will be annihilated. The full scale Monte Carlo simulations  suggest that on average approximately (10-20)\% of the total baryon charge will be lost as a result of 
  traversing of the AQNs through  the Earth's interior, see column for $\Delta B/B$ in table III  in ref.\cite{Lawson:2019cvy}. The average amount of the lost baryon charge depends, of course, on the size distribution and parameter $\alpha$ defined by eq. (\ref{distribution}). To avoid confusion, let us  emphasize one more time   that all these AQNs which get completely annihilated or they lost a finite portion of their baryon charges represent a very tiny portion of all   AQNs during entire evolution of  the Universe as estimated in \cite{Ge:2019voa}.

  Final comment we would like to make  in this subsection is related to the 
    the spallation which represents    a very common  process when   heavy nuclei lose their baryon charge as a result of interaction with a medium. In contrast with conventional nuclei  the spallation cannot play any essential role in AQN survival pattern due to several reasons. First of all, the  gap $\Delta$ in CS phase is typically in 100 MeV range in contrast with conventional nuclear physics where the binding energy normally is  in few MeV range. The most important distinct feature, however,  is as follows.   When a large amount of energy is injected into a heavy nucleus, the spallation takes  place and large number of neutrons may  be liberated leading to the decreasing  of the  baryon charge  of the heavy  original nucleus. Such process cannot occur with AQN because all 
 particles which get excited due to the energy injection  in CS phase  are the coloured objects. Therefore, these elementary excitations  cannot enter   the hadronic vacuum where the normal baryons live and    must stay inside of the  AQN. 
 Therefore, the AQNs do not suffer from spallation processes as heavy nuclei do. This is direct manifestation of the same feature of the AQN construction already mentioned in footnote \ref{first-order} which states that the confined hadronic   and CS phases have different vacua. The same feature precludes transformation of the entire star/planet into a new phase  if the AQNs made of matter stop in the deep interior of the star/planet, see also footnote \ref{matter-AQN}.

  \exclude{
 It should be contrasted with our   present studies related to the neutrino induced neutrons (responsible, according to the proposal (\ref{links}),  for DL annual modulation). In this case    the structure of the nugget's core plays 
the crucial role as neutrino emission originates from within the dense CS matter.  While CS phase realized in QCD for sufficiently  high chemical potential $\mu$  is entirely determined by the SM physics, it nevertheless introduces some new phenomenological parameters which enter the problem.
The corresponding computations have been carried out in previous paper \cite{Lawson:2015cla} for completely different purposes with   different motivation not related to DL modulations. For the convenience of the readers
 we review the relevant features of the CS phase and the main features of neutrino emission  in the Appendix \ref{cfl}.
}

\section{DL modulation by AQNs}\label{AQN-neutrino} 
This section is separated to  four  different subsections, \ref{1}, \ref{2}, \ref{3} and \ref{4} according to four  different elements of the proposal (\ref{links}).

\subsection{AQN flux on Earth}\label{1}
We start  our task with  the {\it   1-st element}  from proposal (\ref{links}) by  estimating   the AQN  hit rate per unit area on earth surface  assuming that  $\rho_{\rm DM}$  is entirely saturated by the nuggets.  The relevant rate has been studied previously in  \cite{Lawson:2019cvy}  for a different problem of computing  the axion flux produced  by the AQNs.     
Now we  estimate the AQN hitting rate   assuming conventional dark matter density    $\rho_{\rm DM}\simeq 0.3  {\rm  {GeV} {cm^{-3}}}$ surrounding the Earth. Assuming the conventional halo model one arrives to the 
  following result 
  \cite{Lawson:2019cvy}:
 \be
\label{eq:D Nflux 3}
\frac{\langle\dot{N}\rangle}{4\pi R_\oplus^2}
&\simeq & \frac{4\cdot 10^{-2}}{\rm km^{2}~yr}
\left(\frac{\rho_{\rm DM}}{0.3{\rm \frac{GeV}{cm^3}}}\right)
 \left(\frac{\langle v_{\rm AQN} \rangle }{220~{\rm \frac{km}{s}}}\right) \left(\frac{10^{25}}{\langle B\rangle}\right).~~~
\ee
The averaging over all types of AQN-trajectories  with different masses $ M_N\simeq m_p|B|$,  with different incident angles and different initial velocities  and different size distribution  does not modify much this estimate.   The result (\ref{eq:D Nflux 3}) suggests that 
the AQNs hit the Earth's surface with a frequency  approximately  once a day   per   $100^2 {\rm km^2}$ area.    The hitting rate for large size objects is suppressed by the distribution function $
 f(B)\propto B^{-\alpha}$ as given by (\ref{distribution}).

 The estimate (\ref{eq:D Nflux 3}) explicitly shows that conventional DM detectors are too small in size to detect AQNs directly as the corresponding flux is many orders of magnitude smaller than the one due to the conventional WIMPs. However, some modern cosmic ray detectors, such as Pierre Auger observatory,  in principle  are capable to study small flux of order (\ref{eq:D Nflux 3}) as suggested in \cite{Lawson:2010uz}. One should also mention that a smaller size IceCube detector  imposes a direct  constraint on  the average baryon charge of the nugget $\la B\ra \geq 3\cdot 10^{24}$, see Appendix A in ref.\cite{Lawson:2019cvy}.

\exclude{In formula (\ref{eq:D Nflux 3}) the average   baryon charge of the nuggets
 $\langle B\rangle\sim 10^{25}$ depends on the nugget's size  distribution   and slightly varies for different models. In eq.  (\ref{eq:D Nflux 3})    we account exclusively the anti-matter nuggets\footnote{we ignore  the matter nuggets which also contribute to $\rho_{\rm DM}$ but  irrelevant for purposes of the present  work} which eventually produce neutrinos as a result of   annihilation processes in the earth's interior.  }
  From (\ref{eq:D Nflux 3}) one can derive the  total hit rate for entire earth's surface which is given by \cite{Lawson:2019cvy}:
\begin{equation}
\label{eq:D Nflux 3 tot}
\begin{aligned}
\langle\dot{N}\rangle
\simeq0.67~ {\rm s^{-1}}
\left(\frac{\rho_{\rm DM}}{0.3{\rm \frac{GeV}{cm^3}}}\right)
 \left(\frac{\langle v_{\rm AQN} \rangle }{220~{\rm \frac{km}{s}}}\right)
 \left(\frac{10^{25}}{\langle B\rangle}\right).
\end{aligned}
\end{equation}
After the nugget hits the surface it  continues to propagate by annihilating the material along its path. The trajectory of the AQN is a straight  line as only small portion of the momentum (and the baryon charge)  will be lost in this journey. The energy produced due to the annihilation events will be isotropically dissipating (in the rest frame of the nugget) along the propagation. 

The rate (\ref{eq:D Nflux 3 tot})  includes   all types of the AQN's    trajectories inside  the earth interior: the trajectories when AQNs hit the surface with incident angle close to $0^o$ (in which  case the AQN crosses the earth core and exits from opposite site of earth) as well as  the trajectories when  AQNs just touch the surface  with incident angle close to $90^o$ (when AQNs leave  the system without much annihilation events in deep underground). 
The result of summation  over all these  trajectories can be expressed in terms of the average mass (energy) loss $\langle \Delta m_{\rm AQN} \rangle $ per AQN.   The same information can be also expressed in terms of the average   baryon charge loss per nugget  $\langle \Delta B\rangle $   as  these two are directly related: $\langle \Delta m_{\rm AQN} \rangle \approx m_p\langle \Delta B\rangle $. The corresponding MC simulations with estimates for $\langle \Delta m_{\rm AQN} \rangle $ have been carried out  in  \cite{Lawson:2019cvy}, see table III in that paper. This information will be  very important for our analysis in next section \ref{2} as it provides a normalization for the total neutrino flux being emitted by the AQNs when they traverse the earth's interior.  

\subsection{Neutrino production from AQNs}\label{2} 

 The {\it 2-nd element } of the proposal (\ref{links}) requires the   estimation of the neutrino intensity  due to the AQN annihilation processes.  
Before we proceed with corresponding  estimates  we want to make a short detour 
related to  the axion production due to the same AQN annihilation events.  We also need to know the basic features of the neutrino spectrum emitted by AQNs. The lessons from that  studies can be used for   estimations of the   neutrino flux from AQNs, which is the main subject of this subsection.  
\subsubsection{Detour on the axion production}
 
 It has been noticed in \cite{Fischer:2018niu} that the large number of axions will be produced  because the axion 
 domain  wall\footnote{As mentioned in Sect. \ref{sec:QNDM} the axion domain wall plays an important  role of a    squeezer stabilizing the nugget. This energy has been accumulated and stored during the QCD epoch at the moment of formation. The corresponding  energy accounts for  a considerable  portion of the nugget's total energy  which is   parametrized by  $\kappa_a\approx 1/3$, see below eq. (\ref{N_a})} will start to shrink during the AQN annihilation events and emit the propagating  axions which can be observed. The corresponding spectrum has been computed in \cite{Liang:2018ecs} where it has been shown that the emitted axions will  have a typical velocities $\la v_a\ra \simeq 0.6 c$ in contrast with conventional galactic axions characterized by small velocities $\sim 10^{-3}c$ such that these two different production mechanisms can be easily discriminated.

The average number of the emitted axions  $\la N_a\ra$ as a result of  the AQN annihilation events   in the earth's  interior can  be   estimated as follows \cite{Liang:2019lya}:
\be
\label{N_a}
\la N_a\ra \approx \kappa_a \frac{\langle \Delta m_{\rm AQN} \rangle}{\la E_a\ra}, 
\ee 
where coefficient $\kappa_a$ determines the relative amount of annihilating energy (per unit baryon charge) transferred to the axion production.  The computation of the coefficient  $ \kappa_a\simeq 1/3$ as well as $\la E_a\ra\simeq 1.3 m_a$ is straightforward  exercise \cite{Liang:2018ecs} as it represents a conventional quantum field theory problem for weakly interacting axion field. 

The energy flux of  the axions  (being averaged over all emission angles and summed over all trajectories)    measured on  earth surface is estimated as  \cite{Liang:2019lya}:
 \be
\label{E_a}
\frac{dE_a}{dt dA} \simeq \kappa_a\la \frac{v_a}{c}\ra \langle\dot{N}\rangle \frac{\langle \Delta m_{\rm AQN} c^2 \rangle}{2\pi R^2_{\oplus}}, 
\ee 
 which has a proper dimensionality ${\rm [GeV\cdot cm^{-2}\cdot s^{-1}] }$ for the energy flux. The axion flux for this mechanism is estimated as
  \be
\label{dN}
\frac{dN_a}{dt dA} \simeq   \kappa_a \la \frac{v_a}{c}\ra\langle\dot{N}\rangle \frac{\langle \Delta m_{\rm AQN}c^2 \rangle}{\la E_a\ra 2\pi R^2_{\oplus}}, 
\ee 
which has a proper dimensionality ${\rm [ cm^{-2}\cdot s^{-1}] }$ for the axion flux.

We are now ready to estimate the neutrino emission from the AQNs,  which is the main topic  of this subsection. One could   follow the same logic of computations as highlighted above, with the only difference is that instead of emission of the axions one should  study the neutrinos,  which are similar to axions
 as they can easily propagate through entire earth interior. This is because the relevant  cross section with  surrounding   material is very small in both cases, and formula (\ref{E_a}) can be applied for  estimation of the neutrino flux on the earth surface with corresponding 
 replacements $\kappa_a\rightarrow \kappa_{\nu}$ and $E_a\rightarrow E_{\nu}$, while  factor $ \langle\dot{N}\rangle$ 
 in eq. (\ref{dN}) of course remains the same. 

 \subsubsection{ On neutrino spectrum  in CS phase}
  
 If we had a  conventional hadronic phase inside the nuggets the corresponding  computations of neutrino emission would be a very simple 
 exercise. Indeed, we know a typical yield of the pseudo Nambu Goldstone (NG) bosons per  annihilation event of a single baryon charge (such as $\bar{p}p$). We also know a typical decay pattern for all NG bosons such as $K \rightarrow \mu\nu$ and  $\pi\rightarrow \mu\nu$ with consequent $\mu$ decays. We also know with very high precision the branching ratios for 
 the non-leptonic  decays of the NG bosons such as $K\rightarrow 3\pi$ and $\eta \rightarrow 3\pi$ with  consequent  decays  to neutrinos. 
 It would allow us to compute the total number of neutrinos  per single  annihilation event. 
    This  would also allow us to compute the energy spectrum of
  neutrinos\footnote{This is precisely the set of assumptions adopted   by ref.\cite{Gorham:2015rfa} where the authors   claimed   that dark matter in the form of AQNs 
cannot account for more than 20$\%$ of the dark matter density. This claim was  based on 
assumption that the annihilation events follow conventional (for confined phase) pattern, in which case 
  a large number of neutrinos  will be produced  in the (20-50) MeV range, 
   where the sensitivity of underground 
neutrino detectors such as Super-Kamiokande have their highest signal-to-noise ratio.   The basic claim of    ref.\cite{Lawson:2015cla} is  that    annihilation processes 
involving an antiquark nugget in CS phase proceed in drastically different way than  assumed in  \cite{Gorham:2015rfa}
when  the lightest pseudo Nambu-Goldstone meson has  mass in the (20-30) MeV range. As a result of this crucial difference the neutrino's energies will be in 15 MeV range, well below the present day constraints.} emitted by the AQNs.

Unfortunately, 
we do not have this luxury of knowing all these key features in the  CS phase.   Therefore, we cannot predict the spectrum of neutrinos emitted by AQNs. The only solid and robust information which is available today is the typical scale of the lightest NG mass
in CS phase, which is normally estimated in $(20-30) {\rm MeV}$ range. This   scale  for NG masses can be  translated to 
 an estimation  for  a typical  neutrino's energy scale   in AQN framework. Assuming that one half  of the lightest NG mass goes to the neutrino's energy we expect that  $\la E_{\nu}\ra \lesssim 15 $ MeV. 
 
 While a typical   energy scale for the NG mass and the corresponding neutrino's energies in the AQN framework is established with a reasonable   accuracy because  it  is based on well established theory of CS phases  \cite{Alford:2007xm,Rajagopal:2000wf},  the computation  of the corresponding neutrino spectrum is a much harder problem.  
 The basic problem is of course, the lack of knowledge of CS phase (in contrast with   the confined phase where all NG masses and relevant branching ratios are measured with very high accuracy). An additional uncertainty also comes from the lack of understanding of the fermion excitations of the  CS phases  which may be, in fact,  the dominant contributors  to  the neutrino fluxes. This is because the NG bosons which are produced as a result of annihilation events in CS phase cannot leave the system and consequently decay to emit neutrino, as they must stay inside  the nuggets\footnote{This is because all excitations in CS phase are the colour objects, and cannot propagate in hadronic vacuum.}. It should be contrasted  with hadronic phase when pions and Kaons (produced as a result of $p\bar{p}$ annihilation) decay to muons and neutrinos in vacuum.   Therefore, the NG bosons in CS phases are likely to be absorbed by fermion excitations (if kinematically allowed), which consequently decay to neutrinos.
 
 Indeed, in unpaired quark matter neutrino emissivity is dominated by the direct Urca processes\footnote{Another direct Urca process $e^-+u\rightarrow  d+ {\nu}_e$ which for antinuggets would correspond to emission of the antineutrino is likely to be strongly suppressed as it requires the presence of the  positrons with sufficiently high energy above the Fermi surface. For low temperature the corresponding  density for positron excitations is exponentially small $\sim \exp(-E_{e^+}/T)$.
 This argument suggests that $\kappa_{\nu}\gg \kappa_{\bar{\nu}}$ as eq. (\ref{N_nu}) states.} such as $d\rightarrow u+e^{-}+\bar{\nu}_e$. In case of antinuggets it should be replaced by anti-quarks with emission of neutrino $\nu_e$ and positron $e^+$ with the energy determined by the energy of the fermion  excitation, which itself assumes the energy of order of the NG mass, see Appendix \ref{nu-antinu} with more comments on this matter.
    
 If this process  indeed becomes the dominant mechanism of the neutrino emission from AQNs than one should expect that 
 the number of neutrinos per single event of annihilation should greatly  exceed the number of anti-neutrino per single event of annihilation, which we assume to be the case. Formally, this case can be expressed  as follows:
 \be
\label{N_nu}
\la N_{\nu}\ra &\approx& \kappa_{\nu} \left[\frac{\langle \Delta m_{\rm AQN} c^2\rangle}{m_pc^2}\right],  ~~
\la N_{\bar{\nu}}\ra \approx \kappa_{\bar{\nu}} \left[\frac{\langle \Delta m_{\rm AQN} c^2\rangle}{m_pc^2}\right], \nonumber \\
 \kappa_{\nu}&\gg& \kappa_{\bar{\nu}}, ~~~~   E_{\nu}  \lesssim  15 ~{\rm MeV}, ~~~~  \kappa_{\nu}\gtrsim 1,
\ee
 where  the coefficients  $ \kappa_{\nu}$ and $\kappa_{\bar{\nu}}$ describe the number of neutrinos and antineutrinos  produced per   single annihilation event, similar to parameter $\kappa_a\simeq 1/3$ entering expression (\ref{N_a}) and  describing  the axion emission   due to the  AQN annihilation events.

We conclude this subsection  with the following generic comment. 
Our system is the strongly coupled gauge theory, the QCD. It should be contrasted with conventional weakly coupled gauge theories when all computations are under complete theoretical control. In our system  
it is very hard to predict  realistic spectra  and intensities  for neutrino and anti-neutrino  fluxes  in the 15 MeV energy range  due to variety of possible phases,  high sensitivity to the parameters   and large number  of possible decay channels    producing  neutrinos    and  antineutrons, as we discussed above. Such  an analysis  could be coined as  ``the nuclear physics of CS phases".  The complexity and uncertainties of such studies (though it is entirely based on the Standard Model physics) is      the main reason  to introduce phenomenological  parameters $\kappa_{\nu}$ and $\kappa_{\bar{\nu}}$ in eq. (\ref{N_nu}) which will be treated in what follows    as  unknown parameters and can only be constrained by experiment.  However, the basic scale of the problem is fixed 
by the lowest NG mass in CS phase with a reasonable accuracy, and it is given by eq. (\ref{N_nu}), i.e. $\la E_{\nu}\ra \lesssim 15 $ MeV. Precisely this basic neutrino energy scale determines the maximum recoil energy around 6 keV  in DL signal.

 \subsubsection{Neutrino flux from AQNs}

 In what follows we need the  expression   for   the neutrino flux similar to our formula for the axion flux (\ref{dN}),
   \be
\label{dN_nu}
\frac{dN_{\nu}}{dt dA} \simeq   \kappa_{\nu} \langle\dot{N}\rangle \frac{\langle \Delta m_{\rm AQN} \rangle}{
 2\pi R^2_{\oplus}m_p}.
\ee 
  Similar expression is also valid for antineutrinos:
  one should replace   $\kappa_{\nu}\rightarrow\kappa_{\bar{\nu}}$ and $dN_{\nu}\rightarrow dN_{{\bar{\nu}}}$ in (\ref{dN_nu}). 
    Now we are ready for the numerical estimates. We use $\langle \Delta m_{\rm AQN} \rangle$ and $ \langle\dot{N}\rangle$ from  subsection \ref{1} to arrive to the following estimate for the neutrino flux in terms of unknown parameter $\kappa_{\nu}$
  \be
\label{nu-flux}
\frac{dN_{\nu}}{dt dA} \simeq   0.6\cdot 10^{6} \cdot  \kappa_{\nu}\cdot \left(\frac{ \langle \Delta B \rangle}{ \la B\ra} \right)\frac{1}{\rm cm^2 \cdot s}, 
\ee 
and similar expression for antineutrinos obtained from  (\ref{nu-flux})  by replacing $\kappa_{\nu}\rightarrow\kappa_{\bar{\nu}}$ and $dN_{\nu}\rightarrow dN_{{\bar{\nu}}}$.
In formula (\ref{nu-flux}) the dimensionless ratio  $\langle \Delta B \rangle/ \la B\ra  $ counts the relative portion  of   baryon charge being annihilated in  the  interior while  AQNs traversing  the earth. This parameter depends on the nugget's size distribution as reviewed in Sect.\ref{sec:Constraints}. Numerically, it is close to $10\%$ for models  with large average charge $\la B\ra\simeq 10^{26}$ and around  $30\%$ for models with smaller  average     charge  $\la B\ra\simeq 10^{25}$ \cite{Lawson:2019cvy}. 
\exclude{\footnote{One should emphasize that the size distribution  is not a new element  here   but  was adopted from solar physics papers devoted to fitting   the solar corona   observations in terms of the so-called nanoflares.   It was  advocated in  \cite{Zhitnitsky:2017rop,Raza:2018gpb}  that  the nanoflares postulated  long ago by Parker can  be identified with the AQN annihilation events  in corona,  such that the AQN size distribution and the nanoflare energy distribution represents one and the same distribution. The computation of the average baryon charge $\la B\ra$  is based  on this identification. It is a highly nontrivial self-consistency check of the entire framework that this value of  $\la B\ra$ is consistent   with available   cosmological, astrophysical, satellite and ground based constraints derived for   fundamentally different physical systems, as reviewed in  section \ref{sec:QNDM}.}, and it is close to $10\%$ for models  with large $\la B\ra\simeq 10^{26}$ and around  $30\%$ for models with smaller   $\la B\ra\simeq 10^{25}$ \cite{Lawson:2019cvy}. 
}
In what follows  to simply things we want to ignore all these numerical factors,  and represent the total neutrino and antineutrino fluxes produced by AQN mechanism  over entire energy range  $0\lesssim E_{\nu, \bar{\nu}}\lesssim 15$ MeV as follows 
  \be
\label{nu-anti-nu}
\frac{dN_{\nu}}{dt dA} \sim    10^{5} \cdot  \kappa_{\nu} \frac{1}{\rm cm^2 \cdot s}, ~~~~~ 
\frac{dN_{\bar{\nu}}}{dt dA} \sim    10^{5} \cdot  \kappa_{\bar{\nu}} \frac{1}{\rm cm^2 \cdot s}.~~~~~
\ee 
 
 It is instructive  to quote few known constraints on neutrino and antineutrino fluxes in this energy band $E_{\nu, \bar{\nu}}\sim 15$ MeV in order to  compare  them with the fluxes produced by the AQN   mechanism as expressed by eq.(\ref{nu-anti-nu}).  The largest flux relevant for this energy band comes from solar $^8B$  which is around  $\Phi_{ {\nu}_e}\simeq 5\cdot 10^6 {\rm (cm^{-2} s^{-1})}$, while solar ``hep" component is close to  $\Phi_{ {\nu}_e}\simeq 8\cdot 10^3 {\rm (cm^{-2} s^{-1})}$ \cite{Bahcall:2004pz}. The atmospheric and diffuse supernova neutrino backgrounds are at least two orders of magnitude smaller than ``hep" component, and can be safely ignored   in our discussions. 

One should also mention the Super-Kamiokande stringent constraint on anti-neutrino flux  $\Phi_{\bar{\nu}_e} < (1.4-1.9){\rm cm^{-2} s^{-1}}$ at large energies  $E >19.3$ MeV 
   \cite{Lunardini:2008xd}     and constraint   $\Phi_{\bar{\nu}_e} < 4\cdot 10^{4}{\rm cm^{-2} s^{-1}}$ at smaller energies $8~ {\rm MeV}<E < 20$ MeV   \cite{Gando:2002ub}. 
  KamLand collaboration \cite{Collaboration:2011jza} reports model-independent upper limit on the anti-neutrino flux  $\Phi_{\bar{\nu}_e} < 10^2{\rm cm^{-2} s^{-1}}$  for every energy bin between 8.3 MeV and 11.3 MeV which  becomes even stronger for higher energies. 
  
  To conclude this subsection we would like to emphasize that the neutrino flux (\ref{nu-anti-nu}) generated by AQNs could be the same order of magnitude as  the dominant $^8B$ solar contributor $\Phi_{ {\nu}_e}\simeq 5\cdot 10^6 {\rm (cm^{-2} s^{-1})}$ in this energy band. It is not presently ruled out by any experiment. The key point here is that the AQN-induced  neutrino flux (\ref{nu-anti-nu}) is the subject of the annual modulation  as it has inherent  DM origin.
  \exclude{\footnote{It is amazing coincidence that these two neutrino fluxes (the AQN-induced and  $^8B$ solar neutrinos which  are originated from drastically different physics, nevertheless could be the same order of magnitude.  The AQN-induced flux is mostly determined by the dark  matter density $\rho_{DM}$ as eq. (\ref{eq:D Nflux 3 tot}) states, in contrast with $^8B$ solar neutrinos determined by internal physics of the sun. A similar ``conspiracy of scales" has been observed previously in our studies  of the so-called ``The Solar Corona Mystery"  \cite{Zhitnitsky:2017rop,Raza:2018gpb} when the Extreme UV   luminosity from solar corona is determined by the dark matter density $\rho_{DM}$ in this framework, and  assumes precisely the observed value without fitting a single parameter.}. }Therefore, it could be the source of the observed DL modulation signal.

 \subsection{Neutrino-induced  Neutrons   }\label{3}
The  {\it   3-rd element}  of proposal (\ref{links}) is the liberation of the neutrons from surrounding rocks due to the coupling with neutrinos.
The intensity of these liberated neutrons is the subject of annual modulation as the corresponding neutron intensity is  proportional to the neutrino flux (\ref{nu-anti-nu}), which itself is directly proportional to the DM flux in form of the AQNs
according to (\ref{eq:D Nflux 3 tot}). 

The idea  that the surrounding neutrons might be the origin of the DL modulations  has been suggested  previously 
in a number of papers, see  \cite{Ralston:2010bd,Nygren:2011xu,Blum:2011jf,Davis:2014cja}. 
This proposal,  of course, remains a subject of debates as the DL collaboration rejects the idea that surrounding neutrons could  play any essential role in their observations \cite{Bernabei:2014tqa}. We will make some comments within AQN scenario 
in next  section \ref{comments} to address this question. 

In the present  subsection we elaborate on the 3-rd element of our   proposal (\ref{links}) by 
      estimating    the neutron flux which is  induced by neutrinos.
      The crucial difference with previous proposals \cite{Ralston:2010bd,Nygren:2011xu,Blum:2011jf,Davis:2014cja} is that the   neutrino-induced neutron flux (\ref{neutrons})
    manifests itself   with  proper    annual DM modulation \cite{Freese:1987wu, Freese:2012xd} characterized by the expected  phase $t_0\simeq 0.4$yr corresponding approximately to June 1 for the Standard Halo Model (SHM).  

We use conventional expressions  from \cite{Davis:2014cja,Bernabei:2014tqa} for cross section $\sigma$ and the number density $n\simeq 10^{29} {\rm m^{-3}}$  of the target  to make straightforward  comparison with  the previous  estimates: 
\be
\label{neutrons}
R_{\nu}^{AQN}\simeq \frac{dN_{\nu}}{dt dA} \sigma (nV)\simeq  10^{-36}  \cdot  \kappa_{\nu} \cdot  (nV) \left[{\rm \frac{neutron}{s}}\right], ~~~~~
\ee
where we used the AQN induced neutrino flux (\ref{nu-anti-nu})  and cross-section $\sigma\simeq 10^{-41} {\rm cm}^{2}$ for the 
neutrino-induced neutron spallation for $^{208}{\rm Pb}$ target.  The effective volume $V$ entering eq. (\ref{neutrons}) will be discussed later in the text.
The AQN-induced neutron rate production per unit volume can be represented 
as follows
\be
\label{neutrons1}
r_{\nu}^{AQN}=\frac{R_{\nu}^{AQN}}{V} \simeq 10^{-2}  \cdot  \kappa_{\nu}~ \left[{\rm \frac{neutron}{day \cdot m^3}}\right]. 
\ee
This rate is approximately one order of magnitude lower if  $\kappa_{\nu}\simeq 1$ in comparison with corresponding estimates adopted in \cite{Davis:2014cja,Bernabei:2014tqa} for the neutron's rate induced by the solar neutrinos. It could be the same order of magnitude as used in \cite{Davis:2014cja,Bernabei:2014tqa}  if $\kappa_{\nu}\simeq 10$, see estimate (\ref{N_nu}).   We return to significance of the estimate (\ref{neutrons1})  later in the text. The only comment we would like to make here is   that the typical kinetic energy of the neutrons liberated by this mechanism will be in the  $10^2$ keV range, see estimate (\ref{E_n}) below.

Indeed, the first thing to notice is that the typical neutrino energy  (\ref{N_nu})   is well above the neutron emission threshold $E_{\nu}> 7.37$ MeV  for $^{208}{\rm Pb}$ target such that neutron spallation is kinematically allowed. Furthermore, 
if one assumes that momentum   resulting from   the neutron spallation 
is mostly transferred to the liberated neutron, such that $ {\bm p}_n\simeq  ( {\bm p}'_{\nu}- {\bm p}_{\nu}) $,
the kinetic energy of the  neutron can be estimated as follows
\be
\label{E_n}
E_n\simeq \frac {{\bm p}^2_n}{2m_n}\sim  10^2~ {\rm keV}.
\ee
One should also add  that there is a sharp  cutoff  of order $\sim 10^2 $ keV   for the neutron's energy (\ref{E_n}) produced by this mechanism.
It is determined by the neutrino energy    (\ref{N_nu}) in 15 MeV range, which itself is  kinematically bound from above  by  
corresponding NG masses  in CS phase as reviewed  in Appendix \ref{cfl}.  The presence of such  sharp cutoff  will be important  element in our arguments supporting the   proposal (\ref{links}).

What do  the neutrons,  characterized by the rate (\ref{neutrons1}) and energy (\ref{E_n}), do 
if they enter the DL detector? This is the subject of the next subsection.

 \subsection{DL modulation (\ref{links}) due to the neutrons}\label{4}
 The  {\it   4-th element}  of proposal (\ref{links}) represents the most controversial  portion of our analysis. 
 We shall try to argue that neutrons characterized by the flux (\ref{neutrons1}) and energy (\ref{E_n}) may serve as  the source of the observed DL annual modulation.  The corresponding computations are very hard to carry out as they are inevitably based on nuclear physics of large number of very  complicated systems. Fortunately, there are many   specific experiments and tests which can be in principle performed,  to be discussed in   sections \ref{comments} and  \ref{tests}. These tests    can support or rule out  the proposal  (\ref{links}). 
 
 The starting point is the standard formula for energy transfer $\Delta E$ as a result of elastic $2\rightarrow 2$ scattering 
 when a target of mass $m_2$ at rest, struck by a particle with mass $m_1$   with kinetic energy $E_n$:
  \be
 \label{delta_E}
 \Delta E=2E_n\frac{m_1m_2}{(m_1+m_2)^2}\left(1-\cos\theta_{\rm CM}\right). 
 \ee 
 The entire section of ref. \cite{Ralston:2010bd} was  devoted for explanation of   numerous uses and  misuses  of this formula in different circumstances. We   agree with most of the comments, careful explanation  of misconceptions, detail analysis of
 neutron-nuclear interaction given by Ralston in   \cite{Ralston:2010bd}. We  defer our specific comments within present context until next section \ref{comments}. 
 
 Now we want to make numerical estimate for the recoil energy    using expression (\ref{delta_E}) and identifying 
 $m_1$ with the neutron characterized by the  kinetic energy (\ref{E_n}), while $m_2\simeq 23m_1$ with the lightest $^{11}$Na nucleon  from DL detector consisting 25 radio-pure  NaI crystal scintillators:
 \be
 \label{E_recoil}
 \Delta E=8.6~ {\rm keV} \left(1-\cos\theta_{\rm CM}\right). 
 \ee  
 The significance of this estimate is hard to overstate  as it unambiguously shows that the recoil energy cannot exceed the value (\ref{E_recoil}) which is amazingly close to 6 keV cutoff observed by DL. Furthermore, the scale  (\ref{E_n}) was not  invented for the specific proposal (\ref{links}), in contrast with many different WIMP-based suggestions  to fit the observed DL modulations.  Rather, this scale  is entirely determined by the cutoff in neutrino energy (\ref{N_nu}) which itself 
 unambiguously fixed by typical NG masses in CS phase. All these scales have been known for quite sometime,  as reviewed in Appendix \ref{cfl}. 
 
 The key element of our proposal (\ref{links}) is that the flux of the  induced neutrons  (\ref{neutrons1}) with kinetic energy (\ref{E_n}), which eventually generates the signal in DL detector with recoil energy (\ref{E_recoil}),   is the subject of annual modulation  because all the intensities are proportional to DM velocity  $\la v_{AQN}\ra$ as eq. (\ref{eq:D Nflux 3 tot}) states. 
Therefore, to describe the corresponding modulations one can use conventional formula  \cite{Freese:1987wu, Freese:2012xd}:
\be
\label{t}
\la v_{AQN}(t)\ra \simeq V_\odot+bV_\oplus\cos\omega(t-t_0)
\ee
where $V_\odot=220~{\rm km~ s^{-1}}$ is the Sun's oribital speed around the galactic center, $V_\oplus=29.8~{\rm km~s^{-1}}$ is the Earth's orbital speed around the Sun, $\omega=2\pi\,\rm yr^{-1}$ is the angular frequency of the annual modulation, and $|b|\leq1$ is the geometrical factor associated with the direction of $\boldsymbol{v}_{AQN}$ relative to the orbital plane of Earth, $t_0\simeq 0.4$yr corresponding to June 1. Hence, it is natural to expect the modulation  must be of order $\mathcal{O}(V_\oplus/V_\odot)\sim10\%$, as incoming flux of the AQN particle  explicitly proportional to $\la v_{AQN}(t) \ra$ according to equation (\ref{eq:D Nflux 3 tot}). The corresponding  value $\langle\dot{N}(t)\rangle$    enters the expression for the AQN-induced neutrino flux (\ref{dN_nu}), eventually generating the neutrons (\ref{neutrons1}).  

The very hard and challenging 
question remains to be answered if this neutron intensity (\ref{neutrons}) is sufficient for explanation of the observed DL modulation.
To  answer  this question one has to   understand  the numerical value for  the effective  volume $V$ entering formula (\ref{neutrons}). 
As we already mentioned this is very complicated  problem of nuclear physics as emphasized and nicely explained  in   \cite{Ralston:2010bd}.

Therefore, instead of theoretical  speculations   about   the value for effective volume entering formula (\ref{neutrons}) we 
reverse the problem and estimate  the volume $V$ which would match the DL modulation.  We suggest several   experiments how this proposal (\ref{links})  and   large rate  (\ref{neutrons}) with effective  volume $V$   can be tested in next two sections \ref{comments} and  \ref{tests}.   

The DL modulation amplitude  in terms of (counts per day) cpd \cite{Bernabei:2018yyw} reads:
\be
\label{observed}
{\rm DL~modulation~ rate}= (0.0103\pm 0.0008)\frac{\rm cpd}{\rm kg ~keV} .~~~~
\ee
This rate must be multiplied to $4~ {\rm keV}$ for (2-6) keV energy range and $250 ~{\rm kg}$ to get total modulation rate
\be
\label{observed1}
{\rm DL~total ~modulation}\simeq  10\left[\frac{\rm counts}{\rm day}\right] . 
\ee
On other hand, assuming that $10 \%$ of neutrons (\ref{neutrons1}) is the subject of annual modulation with proper phase $t_0$
as explained in (\ref{t}) we arrive to the following estimate in terms of the required effective volume $V\equiv L^3$ which saturates the DL modulation:
\be
\label{prediction}
{\rm AQN~induced~modulation} \simeq \kappa_{\nu} \left[\rm \frac{neutrons}{day}\right]   \left(\frac{L}{\rm 10 ~m}\right)^3. ~~~~~~
\ee
  Few comments are in order. First of all, parameter $\kappa_{\nu}$ was introduced as a number of neutrinos being produced as a result of annihilation of a single baryon charge, see  (\ref{N_nu}). It could be as small as  $\kappa_{\nu}\sim 1$, but it could be as large as $\kappa_{\nu}\sim 10$ being consistent with presently available constraints    as mentioned at the end of section \ref{2}. 
  The basic reason  for this uncertainty is that our system is the strongly coupled gauge theory,  to be contrasted with conventional weakly coupled gauge theories when all computations are under complete theoretical control. In our system  
it is very hard to predict  realistic spectra  and intensities  for neutrino and anti-neutrino  fluxes  in the 15 MeV energy range  due to variety of possible phases,  high sensitivity to the parameters   and large number  of possible decay channels    producing  neutrinos    and  antineutrons, as we discussed above. This deficiency in our computational power should not be treated as the weakness of the proposal. Instead, 
it should be considered as a consequence of the complexity of the system.  
  
  The observed rate (\ref{observed1}) matches the  AQN induced modulation (\ref{prediction}) if $\kappa_{\nu}\simeq 10$ and $L\sim 10$ m. This required length $L\sim 10$ is definitely much greater   than the neutron's mean free path $\lambda\simeq 2.6$ m which was extracted  from the  studies on the muon-induced background  \cite{Barker:2012nb} and  adopted  by refs. \cite{Davis:2014cja,Bernabei:2014tqa} in the context of the present work on DL modulations.     In next section \ref{comments} we  comment 
  on the  consistency (\ref{prediction}) with DL observations, while in section \ref{tests}  we make few comments on relation to other experiments. In the same section \ref{tests} we also suggest   possible tests (such as the measuring of the spatial directions of moving neutrons along with time modulation) which could  support or rule out the proposal (\ref{links}). 
 
  \section{Comments on DL arguments.}\label{comments}
  The DL collaboration, of course, discussed  the possibility that their signal is associated with neutron flux (induced by muons or neutrinos or both).  In fact, entire paper \cite{Bernabei:2014tqa} was devoted to analysis of a possible role of neutrino-induced  and muon- induced neutrons. These possibilities were discarded in  \cite{Bernabei:2013xsa,Bernabei:2014tqa,Bernabei:2016bkl,Bernabei:2018yyw} based on the following arguments:
  
  1. Modulation phase arguments;
  
  2. Energy range arguments;
  
  3. Intensity arguments.
  
  We want to make few comments on each of the items from this list. We start with the modulation arguments, while the energy   and intensity arguments will follow.
  
1.   Quote from \cite{Bernabei:2014tqa}: `` ... It is worth noting that neutrons, muons and solar neutrinos are not competing  background when DM annual modulation signature is investigated since in no case they can mimic this signature...". In the proposed scenario (\ref{links}) 
  this argument obviously does not apply because both the neutrino flux (\ref{nu-anti-nu}) and the neutron flux (\ref{neutrons}) with typical energy (\ref{E_n}) are automatically the subject of the annual modulation (\ref{t}) with proper phase $t_0\simeq 0.4$yr corresponding to June 1. This is because   the source of the modulation in this framework has truly genuine DM origin represented by AQNs.  
  
 2. The DL has carried out the  comprehensive studies  on dependence of the annual  modulation as a function of the   energy interval. 
 The claim is that the modulation is not observed above the energy $\sim $ 6 keV. In particular  the modulation amplitude for the energy above 6 keV for the whole data sets (DAMA/NaI, DAMA/LIBRA-phase-1, DAMA/LIBRA-phase-2)  is shown to be consistent with zero, see Fig 11 in ref. \cite{Bernabei:2018yyw}. This property is indeed very hard to understand  in terms of the conventional physics advocated in \cite{Ralston:2010bd,Nygren:2011xu,Blum:2011jf,Davis:2014cja}. 
   
 At the same time this unique feature of the system (characterized by  a sharp cutoff   at  $\sim $ 6 keV)  automatically emerges  in our framework. Indeed,   the neutrino flux (\ref{nu-anti-nu}) with typical neutrino energy $E_{\nu}  \sim  15 ~{\rm MeV}$ is determined in our framework by the NG masses in the CS phase, see (\ref{kappa}). The corresponding neutrino-induced neutron flux (\ref{neutrons}) is characterized by the  typical energy (\ref{E_n}) formulated in terms of $E_{\nu}$. 
 The sharp cutoff for the recoil energy (\ref{E_recoil}) in this framework (which   falls into the proper   $\sim$ 6 keV energy range) 
  is   determined by (\ref{E_n}) which essentially determined by   the NG masses in CS phase as reviewed  in Appendix \ref{cfl}. One should emphasize that all these energy scales have not been specifically invented  in this work to explain the observed DL modulations with (1-6) keV energy;  rather the relevant energy scales have  been established long ago in unrelated studies for different purposes in a different context.   
  
  3. The neutrino flux (\ref{nu-anti-nu})  originated from AQNs in this framework is  lower than the background  solar neutrino flux $\Phi_{ {\nu}_e}\simeq 5\cdot 10^6~ {\rm cm^{-2} s^{-1}}$ for 
  this energy band (originating  from solar $^8B$) at least by factor of 5 for $\kappa_{\nu}\simeq 10$, as mentioned at the end of section \ref{2}.
  The corresponding neutrino-induced neutron flux (\ref{neutrons}) is also must be lower in comparison with intensity of neutrons induced by the solar neutrinos. However, the key point is that this {\it subdominant  neutron component} is originated from AQNs, and therefore, is the subject of conventional DM annual modulation (\ref{t}) with proper phase $t_0$.  
  
  Is the corresponding neutrino-induced  neutron's intensity is sufficient\footnote{\label{matches} As we already mentioned above, the question is essentially reduced to the quantitative understanding  of the effective volume $V=L^3$ which enters (\ref{prediction}). 
Our estimates from previous section show that if $L\simeq 10$m, the AQN induced rate (\ref{prediction})  matches the observed rate (\ref{observed1}) with $\kappa_{\nu}\simeq 10$.} to explain  DL modulation? 
  The DL argued that the answer is ``No" \cite{Bernabei:2013xsa,Bernabei:2014tqa,Bernabei:2016bkl,Bernabei:2018yyw}. However, the DL      arguments   on the neutron's intensity  were  challenged in \cite{Ralston:2010bd}.  We have nothing new to add to these extensive discussions on possible role of  neutrino-induced neutrons. Instead of speculations about  this very complex nuclear physics system with complicated resonance structure, we suggest to test the proposal (\ref{links}) by measuring the modulation, intensity  and the {\it directionality}  in coordinate space of the neutrons which are responsible for the recoil (\ref{E_recoil}).  
  
  The subdominant flux of the AQN-induced neutrons can be, in principle,  discriminated from the dominant components, including the solar neutrino-induced neutrons if the {\it direction} of the neutron's momentum and the modulation are measured. This is because the solar neutrinos  are propagating from a single direction in the sky, while AQN-induced neutrinos   (which have  truly genuine DM origin) are randomly distributed in space. 
  This topic on possible tests of the proposal (\ref{links}) represents the subject of   the next section \ref{tests}.
  
 \section{Relation to other experiments.  Possible  future tests.}\label{tests}
 This section is separated to three different subsections.  First, in subsection \ref{previous} we make few comments on  the previous experiments 
 which exclude the  DL  signal if
 interpreted  in terms of WIMP-nuclei interactions. We continue with more recent analysis in subsection \ref{recent} where we make  few comments on some recent experiments  designed  to
 reproduce (or rule out)  the DL signals. Finally, in subsections \ref{future}  and \ref{correlations} we offer few novel possible tests which can support or refute  the proposal (\ref{links}) explaining the DL signal  in terms of the AQN-induced neutrons. In particular, in subsection
 \ref{correlations}

 \subsection{Previous experiments}\label{previous}
  We start with 
   few comments on the experiments which exclude  the DL results.   The corresponding  collaborations  include but not limited to: 
 CDEX \cite{Yue:2014qdu}, CDMS- II \cite{Ahmed:2010wy,Ahmed:2012vq}, EDELWEISS-II \cite{Armengaud:2012pfa}, LUX \cite{Akerib:2013tjd}, SuperCDMS \cite{Agnese:2014aze}, XENON10 \cite{Angle:2011th} and XENON100 \cite{Aprile:2012nq}, CoGeNT \cite{Aalseth:2014eft}. The main claim of these collaborations can be formulated as follows:  if DL modulation is interpreted in terms of WIMP-nuclei interactions with given $\sigma$ and given $m_{WIMP}$ then DL signal is excluded with very high level of confidence \cite{Tanabashi:2018oca}.  
 
 From the perspective of the proposal (\ref{links}) it could be a number of reasons of why DL observes the signal while other collaborations do not. First of all, most of the collaborations (with few exceptions such as  CoGeNT \cite{Aalseth:2014eft} and CDMS- II \cite{Ahmed:2012vq}) did not carry out some dedicated   studies on the  time modulation, which was the crucial ingredient  in DL arguments. From the AQN perspective the time modulation  is the key element when  the subdominant neutron flux  can  manifest   itself  if proper time modulation studies are performed. 
 
 Another reason (for negative results) could be related to  different neutron shields used by different collaborations.   We refer to paper by Ralston \cite{Ralston:2010bd} where  the subject on   complex behaviour of neutrons  in complicated environment is nicely presented. This analysis  obviously shows that even minor differences in neutron shields may have dramatic effects and drastically change the impact of neutron's background.    
 
 Yet, one more reason, probably the most important one  in context of the present work,  is as follows. The AQN-induced   neutrons with energy (\ref{E_n})  are scattering off Na in DL experiment generating recoil energies which fall into the (2-6) keV bin according to (\ref{E_recoil}).
 The recoil energies for the  heavier targets such as xenon or germanium in different experiments could be below threshold because  the energies of the time modulated neutrons 
 (\ref{E_n}) are bound from above with a cutoff  being  determined  by energies  of the AQN-induced neutrinos. The energies of these neutrinos are also bound from above and    cannot exceed (\ref{N_nu}) as they are  determined by NG masses in CS phase.  As we already mentioned previously, 
 all these energy scales in proposal (\ref{links}) have not been invented to fit the DL signals. Rather the relevant energy scales have been established long ago in unrelated studies for different purposes in a different context. 
 
 \subsection{Recent activities }\label{recent}
 Now we want to make few comments on recent dedicated experiments which were specifically designed to test DL annual modulation signal. It includes COSINE-100   \cite{Adhikari:2018ljm,Adhikari:2018fmv,Adhikari:2019off}, ANAIS-112 \cite{Amare:2019jul} and   CYGNO
  \cite{CYGNO:2019aqp}   experiments. We also want to mention   other type of  experiments \cite{Capparelli:2014lua,Hochberg:2016ntt,Cavoto:2017otc} which were not originally designed to test DL annual modulation signal.
   However, their  capabilities to measure the directionality could play a decisive role  in detecting of the DM signals. 
    To be more specific, we choose to mention these     experiments due to the following reasons:
 
 The  COSINE-100  experiment was mostly motivated by DL annual modulation.  The aim of the collaboration is  to reproduce (or refute)  the signal and to search for possible origin for the  modulation, if observed. The COSINE-100 collaboration uses the same target medium (sodium iodide) which is the crucial element in the context  of the present proposal (\ref{links}) as recoil energies  fall into the (2-6) keV bin in our scenario according to (\ref{E_recoil}). Presently the COSINE-100 data is consistent with both a null hypothesis and DL (2-6) keV  best fit value with 68\% confidence level. More data are obviously needed.  It is important  that COSINE-100 is planning to measure the neutron's intensity and neutron's modulation \cite{Adhikari:2018fmv}, in which case the  COSINE-100 would know if the possible modulation is due to the neutrons\footnote{\label{COSINE}I thank G. Adhikari for answering a  large number of my questions during the PATRAS-2019 axion meeting   about the future plans of the collaboration.}. It is obviously the key element of the  proposal (\ref{links}) based on the subdominant AQN-induced  component  of neutrons which, however, manifests itself by annual modulation.  
 
 The ANAIS-112 collaboration also uses the same target medium  as DL and the COSINE-100. The ANAIS-112 has recently published the first results on annual modulation \cite{Amare:2019jul}. Their best fits are incompatible at $2.5\sigma$ with DL signal. The goal is to reach the sensitivity at  $3\sigma$ level in five years. As the ANAIS-112 collaboration uses the same target material our comments from the previous paragraph in context  of the present proposal (\ref{links}) also apply  to ANAIS-112 experiment especially as ANAIS-112 and  COSINE-100 agreed to combine their  data. 
 
 The  CYGNO proposal  \cite{CYGNO:2019aqp} is different from  the  COSINE-100 and the ANAIS-112  experiments due to the capability    to measure the {\it directionality}
 which is the key element of the CYGNO  \cite{CYGNO:2019aqp} proposal. It is important that it will be located at the same site (LNGS) where DL is located. 
 Therefore, the neutron flux must be the same, including the subdominant AQN-induced component (\ref{prediction}) which is the subject of annual modulation. The collaboration is planning to   measure (initially) the neutron flux and its modulation  without neutron shielding\footnote{\label{CIGNUS}I thank Elisabetta Baracchini  for answering a  large number of my questions during the PATRAS-2019 axion meeting   about the future plans of the collaboration.}.  Such measurements may play a key role in supporting   (or ruling out) the   proposal (\ref{links})  because the AQN-induced neutrons are responsible for the recoil (\ref{E_recoil}).  The collaboration is also planning in future to reach the neutrino floor by measuring the neutrinos and  their directions. In particular the    CYGNO  could discriminate the neutrinos from the sun  by identifying their  directions. Furthermore, the CYGNO instrument  will be capable to determine nuclear recoil directions, which would allow the collaboration  to discriminate WIMP-like DM from AQN-induced events   (\ref{links}).  
   Such measurements, if successful,  would obviously play  an important  role in supporting (or ruling out) the   proposal (\ref{links}).
   In addition to that,   the dominant solar  neutrinos in the energy range
    $E_{\nu} \lesssim 12 $ MeV could be discriminated from subdominant
   AQN-induced neutrinos (\ref{kappa}). Furthermore,  neutrinos in the energy band $E_{\nu} \geq 12 $ MeV cannot be originated from the Sun at all as a result of $^8B$ threshold. The discovery of such neutrinos and measuring of their annual modulation with intensity in the range (\ref{nu-anti-nu}) would be enormous support for the proposal (\ref{links}) as the flux  of the   atmospheric   neutrino  is at least three orders of magnitude lower than  (\ref{nu-anti-nu}).
   
  We also want to mention  several  other  experiments   \cite{Capparelli:2014lua,Hochberg:2016ntt,Cavoto:2017otc}  which are capable to measure the directionality.  The  idea is to use  the  carbon nanotube arrays or the graphene layers to measure the directionality of the DM signals.  As we mentioned above, the measurements of the directionality could play a decisive role  in detecting of the DM signal. 
 
 \subsection{Possible future  tests}\label{future}

 We already mentioned  in  previous subsection \ref{recent} few possible tests which can support or rule out the proposal (\ref{links})
 with existing or planning experiments: COSINE-100   \cite{Adhikari:2018ljm}, ANAIS-112 \cite{Amare:2019jul}, CYGNO 
  \cite{CYGNO:2019aqp}. 
 In this subsection we want to mention a specific for the AQN framework phenomenon 
 when the intensity of the AQN-induced neutrinos may be amplified by very large factor (up to $10^4$) which greatly increases the chance for discovery
such  AQN-induced neutrons which always accompany the neutrinos according to section \ref{3}. Therefore, this neutrino amplification factor will be obviously accompanied  by  the corresponding amplification of the neutrino- induced neutron flux     (\ref{neutrons}) and  (\ref{neutrons1}).

The idea was originally formulated  for the axions in \cite{Liang:2019lya}, and the effect was coined as the ``local flash". The computations can be easily generalized for the neutrinos in straightforward way,   see Appendix \ref{local flashes} with some  technical details. It can be explained as follows.

If the AQN hits the surface at distance $d\ll R_\oplus$ from the detector the short-lasting flash occurs
with  amplification factor  $A_{\nu}(d)$ measuring the relative short lasting spark  of the neutrino flux 
with respect to the   neutrino flux (\ref{nu-anti-nu}) computed by averaging  over entire earth surface over long period of time. 
The amplification   $A_{\nu}(d)$ is highly sensitive to distance $d$ and can be approximated as follows, see (\ref{nu3}) for derivation:
\begin{equation}
\label{nu-amplification}
A_{\nu}(d) 
\simeq\frac{1}{\langle \dot{N}\rangle\langle \Delta t\rangle}
\left(\frac{R_\oplus}{d}\right)^2\ .
\end{equation}
One should note that the correction to the neutrino flux (\ref{nu2}) due to the traversing of a nearby AQN
depends on unknown parameter $\kappa_{\nu}$.
However, the relative amplification $A_{\nu}(d)$ with respect to the averaged   neutrino flux (\ref{nu-anti-nu}) does not depend on $\kappa_{\nu}$
as eq. (\ref{nu-amplification}) states. 
In   formula  (\ref{nu-amplification})
 $\langle \dot{N}\rangle$ is determined by eq. (\ref{eq:D Nflux 3 tot}),  while $\langle\Delta t\rangle \simeq  2R_\oplus/\la v_{\rm AQN}\ra$
is the  time for the AQN to cross the earth averaged over entire ensemble of AQN's trajectories traversing the earth. 

As one can see from (\ref{nu-amplification}) a huge amplification may indeed occur  for $d\ll R_\oplus$.  
However, the probability for such event to happen is very tiny, and can be estimated as 
\be
\label{nu-Event rate}
{\rm Event~rate}
\simeq\frac{1}{[\langle\dot{N}\rangle\langle\Delta t\rangle^3]^{\frac{1}{2}}}\cdot\frac{1}{A_{\nu}^{\frac{3}{2}}},
 \ee
see Appendix \ref{local flashes} with details. The ``local flash" lasts for a short period of time which can be estimated as follows
\begin{equation}
\label{nu-tau}
\tau
\simeq\frac{2d}{\la v_{\rm AQN}\ra} 
\simeq\left(\frac{\langle\Delta t\rangle}{\langle\dot{N}\rangle}\right)^{1/2}
\frac{1}{A^{\frac{1}{2}}_{\nu}}.
\end{equation}
We summarize in Table \ref{tab:local flashes}   few choices of time duration $\tau$ and  the event rate as a function of amplification factor $A_{\nu}$. In particular, it would be a daily short  lasting  ``flash" when the intensity of the subdominant AQN-induced neutrino component (\ref{nu-anti-nu})  is amplified by factor $A_{\nu}(d)\simeq 10^2$ such that  it becomes the dominant one for a short period of time  lasting for about 1 second. 
\begin{table} 
\captionsetup{justification=raggedright}
	\caption{Estimations of Local flashes for different $A_{\nu}$, adopted from \cite{Liang:2019lya}: the time duration, and the corresponding event rate. } 
	\centering 
	\begin{tabular}{ccc}
		\hline \hline
		$A_{\nu}$ &  $\tau$ (time span) & event rate \\ 
		\hline 
		1 & 10 s & 0.3 $\rm min^{-1}$ \\ 
		$10$ & 3 s & 0.5 $\rm hr^{-1}$ \\ 
		$10^2$ & 1 s & 0.4 $\rm day^{-1}$ \\ 
		$10^3$ & 0.3 s & 5 $\rm yr^{-1}$ \\  
		$10^4$ & 0.1 s & 0.2 $\rm yr^{-1}$ \\  
		\hline 
	\end{tabular}
	\label{tab:local flashes}
\end{table}

Important lesson to be learnt from these estimates is as follows. A subdominant neutrino flux induced by AQNs 
may become a dominant portion of the neutrinos, overpassing the solar neutrino flux in this energy band for very short period of time.  Needless to say, that this AQN-induced neutrino flux (\ref{nu-anti-nu}) and the corresponding neutron flux  (\ref{neutrons})     are also the subject of annual (\ref{t}) and daily modulations\footnote{Daily modulations with intensity around (1-10)\%   are similar in magnitude as annual modulations.   These daily modulations are  unique for this type of DM, and not shared by any other DM models, see \cite{Liang:2019lya} for the details.}, similar to the ones studied for the axion search experiments \cite{Liang:2019lya}. 
The measure of directionality and modulation as described  in previous subsection \ref{recent} may help to discriminate this subdominant AQN-induced neutrino flux from the dominant solar $^8B$ component.   

\subsection{Search for correlations with other AQN-induced phenomena}\label{correlations}
The flux  (\ref{eq:D Nflux 3}) suggests that 
the AQNs hit the Earth's surface with a frequency  approximately  once a day   per   $100^2 {\rm km^2}$ area, which is precisely the source for a short-lasting amplification in the neutrino production discussed  in previous subsection \ref{future}. It is important to emphasize that such events  occur  along with  other processes which   always   accompany  the AQNs  propagation  through the atmosphere and the Earth's underground. Therefore, there will be always some correlations between the amplifications in the neutrino flux and other associated  phenomena in the vicinity  of the area where   AQN event occurs. 

For example, 
 if the DM detector (sensitive to neutrino-induced neutrons such as DL)   and an axion search detector   are   localized close to each other on a  distance $d_{\nu  a}\sim d$  there will be  the axion signal due to the amplification $A_a$ and the neutrino-induced neutron  signal due to the amplification $A_{\nu}$. These signals  must be correlated in form of two almost simultaneous short lasting sparks  between these two  signals in two different detectors.    The  observation  of these cross-correlated signals  (collected during a long period of time and by measuring the directionality in DM detector to discriminate the background) would unambiguously  support the proposal (\ref{links}) on the nature of the observed DL modulation signal.  
 Similar cross-correlation between different synchronized axion stations from a Global Network as suggested in \cite{Budker:2019zka} and a nearby  neutrino detector would also strongly   support the proposal (\ref{links}).

In particular, the position of the Center for Axion and Precision Physics Research located at  Daejeon and COSINE -100 detector 
located at the Yangyang Underground Laboratory in South Korea obviously satisfy the criterial $d_{\nu  a}\ll 0.1 R_{\oplus} $
when strongly    correlated amplifications for $A_{\nu}$ and $A_a$  may occur in both detectors almost simultaneously with time delay of order of few seconds.  
 
Another correlation which is worthwhile to  study can be explained as follows. It has been recently argued that the AQN propagating in the Earth's atmosphere and underground   emits the infrasound and weak seismic waves \cite{Budker:2020mqk}. In fact one such event, according to  \cite{Budker:2020mqk},   occurred on July 31-st 2008. It   was properly recorded by the dedicated Elginfield Infrasound Array (ELFO) near London, Ontario, Canada and corresponds to relatively large  nugget\footnote{The frequency of appearance for such large   nuggets is very tiny  according to (\ref{distribution}). This explains why such events occur approximtaely once every 10 years instead of observations them once a day  which would be the case for  much more common nuggets with $B\sim 10^{25}$.} with $B\simeq 10^{27}$ if interpreted as AQN event \cite{Budker:2020mqk}.    The infrasound detection was accompanied by non-observation of any meteors by an all-sky camera network. The impulses were also observed seismically as ground coupled acoustic waves around South Western Ontario and Northern Michigan. The estimates  \cite{Budker:2020mqk} for the infrasonic frequency
 $\nu\simeq 5$ Hz and overpressure $\delta p\sim 0.3$ Pa are consistent with the ELFO record. It has been also proposed   in  \cite{Budker:2020mqk} a detection strategy for a systematic  study to search for  such   events  originating from much smaller and much more common AQNs  with typical $B\simeq 10^{25}$ by using Distributed Acoustic Sensing (DAS). 
 
 Our original remark here is that the amplification in the neutrino flux (and corresponding enhancement in the neutrino-induced neutrons)   as discussed in previous subsection \ref{future} must be accompanied by  infrasonic  and weak seismic waves 
 which can be  studied by the DAS instruments     implemented in networks of optical-fiber telecommunication cables.     The observation for  such correlations,  
 if successful, is  absolutely unique to AQN framework and  would obviously play an important role in supporting of  the proposal. 
 
 In particular, the DL  (and future  CYGNO detector) is located at the LNGS site with large number of    seismic detectors located in the same area. Proposal is to search for the correlations between enhanced flux of neutrons and weak seismic and
  infrasound events with delay  measured in fraction of a  second depending on a precise    localization of the seismic detectors.  The measurements of the directionality by CYGNO would be the key element in establishing such a correlation.

 \section{Concluding Comments}  
 The main results of our work can be summarized as follows:
 
 1.We argued that the annual modulation observed by DL might be explained as a result of the AQN-induced neutrons through the chain 
 (\ref{links}).   In this framework  the annual modulation    has  truly genuine DM origin, though it is manifested indirectly. 
 
 2.We also argued that the   recoil  energy must have a sharp cutoff at   $\sim 6~ {\rm keV}$  consistent with the  observed DL signal.   
 
 3.  We proposed specific tests which can support or rule our the proposal (\ref{links}),  see subsection \ref{future}. 
 
 4. We proposed to study a specific correlation between the amplification in the AQN-induced neutron  flux  and the impulses of  the infrasound and weak seismic waves, which if found, would strongly support our proposal, see subsection \ref{correlations}. 
 
 Why should we consider this AQN model seriously? There is a number of reasons. Originally, this model  was invented to  explain 
 the observed relation $\Omega_{\rm DM}\sim \Omega_{\rm visible}$ and the baryon asymmetry  of the Universe as two sides of the same coin, 
when  the   baryogenesis framework   is replaced by a   ``charge separation" framework,  as reviewed  in Section \ref{sec:QNDM}. After many years   since its original formulation  this model remains to be  consistent with all available cosmological, astrophysical, satellite and ground based constraints, where AQNs could leave a detectable electromagnetic signature.  Furthermore, it is shown that the AQNs   can be formed and  can  survive the  unfriendly environment during the evolution of the early Universe, such that  they entitled to  serve as the DM candidates.   Finally, the same AQN framework    may also explain a number of other (naively unrelated) observed phenomena such as excess of the galactic diffuse emission in different frequency bands, the so-called ``Primordial Lithium Puzzle" and ``The Solar Corona Mystery", the seasonal variations observed by XMM-Newton observatory,  to name just a few, see Section \ref{introduction} for the references.

We want to emphasize that 
all  these cosmological puzzles mentioned in Section \ref{introduction} 
 could  be resolved within the AQN framework with the {\it same set of physical parameters} being used  in the present work on explanation of DL modulation signal. 
 
 The observation of the subdominant AQN-induced neutrons   by measuring the directionality and modulation as discussed in Section \ref{recent} would be   a direct manifestation of the AQN dark matter model.  The observation of variety of amplifications    as discussed in Section \ref{future} would be also a strong support for the proposal (\ref{links}). Finally, the recording of  the  correlations between the AQN induced neutrino amplification and the impulses of  the infrasound and weak seismic waves would strongly support our proposal as argued in subsection \ref{correlations}. 
 We finish this work on this positive and optimistic note.

\section*{Acknowledgements}
I am thankful to  G. Adhikari for answering my questions on present status and    future plans for COSINE-100 collaboration, see footnote \ref{COSINE}, and  
Elisabetta Baracchini for elaboration on  future plans for CYGNO  collaboration, see footnote \ref{CIGNUS}
during the PATRAS-2019 axion  meeting in Freiburg in June 2019.  
I am thankful to Maria Martinez [ANAIS-112] and  Elisabetta Baracchini [CYGNO] for correspondence. 
I am also thankful to  Hyunsu Lee [COSINE] and Maria Luisa Sarsa [ANAIS-112] for discussions during the 
``Conference on Dark World" in Daejeon, Korea in  November  2019, where this work was presented. 
This research was supported in part by the Natural Sciences and Engineering 
Research Council of Canada.

 \appendix
 \section{Neutrino spectrum from the Axion Quark Nuggets}\label{cfl}
 The main  goal of this Appendix is to give a short overview of the basic results from ref.\cite{Lawson:2015cla}
 regarding  the neutrinos    emitted by  AQNs captured by  the Sun. The paper \cite{Lawson:2015cla} was written in response 
 to  the claim made in  \cite{Gorham:2015rfa} that dark matter in the form of AQNs 
cannot account for more than 20$\%$ of the dark matter density. This claim was  based on 
constraints on the neutrino flux in the (20-50) MeV range where the sensitivity of underground 
neutrino detectors such as Super-Kamiokande have their highest signal-to-noise ratio. 

However, the estimates  \cite{Gorham:2015rfa} were based on an assumption that the annihilation processes between antimatter from AQNs and normal   material from the Sun has the same spectral features as conventional baryon- antibaryon annihilations which  typically produce a large number of 
pions which eventually decay though an intermediate muon and thus generate a 
significant number of neutrinos and antineutrinos in the the (20-50) MeV range.   

However, as it has been  argued in  \cite{Lawson:2015cla}    the critical difference in the case of  annihilation processes 
involving an antiquark nugget is that the annihilation proceeds within the 
colour superconducting (CS) phase where the energetics are drastically  different. 
The main point is that in most CS phases the lightest pseudo Goldstone mesons 
(the pions and Kaons) have masses in the  20 MeV range \cite{Alford:2007xm,Rajagopal:2000wf} 
in huge contrast with the hadronic confined phase where $m_{\pi}\sim 140$ MeV.   As a result of this crucial difference the decay of light pseudo Goldstone mesons of the CS phase cannot produce  
neutrinos in the 20-50 MeV energy range and are not subject to the 
SuperK constraints employed in \cite{Gorham:2015rfa}. Instead, the pseudo Goldstone mesons of the CS phase
produce neutrinos in  15 MeV range. 

These unique spectral features of the neutrinos  emitted by AQNs  play the  key role in our proposal  
suggesting  that the observed cutoff in DL modulations at 6 keV is ultimately related to the cutoff in the neutrinos energies at  15 MeV   emitted by AQNs.   The emergence of this  new  15 MeV energy scale  is the subject of the next subsections.

 \subsection{ Nambu-Goldstone modes in CS phase}\label{sec:NG}
  
  There are many possible CS phases due to the generation of a gap $\Delta$ through 
different channels with slightly different properties.
While the relevant physics is a part of the standard model, QCD with no free parameters, 
the corresponding phase diagram is still a matter of debate as it strongly depends on 
the precise numerical value of the gap $\Delta$, see review articles 
\cite{Alford:2007xm,Rajagopal:2000wf}.
For our purposes though the key characteristics are very much the same for all
phases. Therefore, we limit ourself below to reviewing the most developed, 
the so called CFL (colour flavour locking) phase.  
The spontaneous
breaking of chiral symmetry in colour-superconductors gives rise to
low-energy pseudo--Nambu-Goldstone (NG) modes with similar quantum
numbers to the mesons (pions, kaons, etc.).  These objects, however, are
collective excitations of the CS state rather
than vacuum excitations as is the case for the conventional confined hadronic phase.
The finite quark masses explicitly break chiral symmetry, giving rise to these
``pseudo''--Nambu-Goldstone modes on the order of  20 MeV, in huge contrast 
with the hadronic confined phase where the lightest mass meson has 
$m_{\pi}\simeq 140$ MeV.  
  
To be more precise, we consider large $\mu$ limit for which the masses and 
other relevant parameters in the CFL phase can be explicitly computed 
\cite{Alford:2007xm,Rajagopal:2000wf}: 
\be
\label{NG}
m^2_{\pi^{\pm}}&\simeq&\frac{2\bar{c}}{f_{\pi}^2}m_s(m_u+m_d), ~~ f_{\pi}^2\sim \mu^2\nonumber\\
m^2_{K^{\pm}}&\simeq&\frac{2\bar{c}}{f_{\pi}^2}m_d(m_u+m_s) , ~~~ \bar{c}\simeq \frac{3\Delta^2}{2\pi^2} \nonumber\\
m^2_{K^0}&\simeq&\frac{2\bar{c}}{f_{\pi}^2}m_u(m_d+m_s) .
\ee
  As one can see from (\ref{NG}) 
 the NG bosons are much lighter than in vacuum. This is because their 
masses are proportional to $m_q^2$ rather than to $m_q$, as at zero chemical potential. 
As a result, the lightest NG meson, the kaon, has a mass in the range of 10 to 20 MeV 
depending on precise value of $\Delta$ and  $\mu $ MeV, see 
for example \cite{Rajagopal:2000wf}.  
  
Another important difference between the NG modes in dense matter and in vacuum
is in the dispersion relations for the NG modes which assume the following form, 
see for example \cite{Alford:2007xm}:
\be
\label{NG1}
E_{K^{\pm}}&=&\mp\frac{m_s^2}{2\mu}+\sqrt{v_{NG}^2 p^2+m^2_{K^{\pm}}}\\
E_{K^{0}}&=&-\frac{m_s^2}{2\mu}+\sqrt{v_{NG}^2 p^2+m^2_{K^{0}}} \nonumber\\
E_{\bar{K}^{0}}&=&+\frac{m_s^2}{2\mu}+\sqrt{v_{NG}^2 p^2+m^2_{K^{0}}} \nonumber\\
E_{\pi^{\pm}}&=& \sqrt{v_{NG}^2 p^2+m^2_{\pi^{\pm}}}, ~~~~~  v_{NG}^2=\frac{c^2}{3}.  \nonumber
\ee
such that the rest energy of the lightest NG mesons does not exceed 
the 10-20  MeV range. In fact $E_{{K}^{0}}$ may even vanish, in which 
case the ${K}^{0}$ field forms a condensate (the so-called $CFL~ K^0$-phase). In these formula   $v_{NG}$ deviates from speed of light $c$ due to the 
explicit violation of the Lorentz invariance in the system
such that the dispersion relations for all quasiparticles are drastically different 
from their counterparts in vacuum.

One should comment here that the dispersion relations for the NG modes within the  
anti-nuggets (which is most relevant for our purposes) can be obtained from (\ref{NG1}) by replacing  $\mu\rightarrow  -\mu$ 
such that the lightest NG states become the $\bar{K}^{0}$ and $K^-$ for  
nuggets made of antimatter. This comment is important  for identification of the neutrino and  anti-neutrino spectra to be discussed in next subsection \ref{neutrino}.

 \subsection{Neutrino emission from  NG bosons}\label{neutrino}
The neutrino emission from CFL phase quark matter has been studied previously 
in a number of papers mostly in context of the physics of neutron stars, see the 
original papers \cite{Jaikumar:2002vg,Reddy:2002xc,Reddy:2003ap} and the review 
article \cite{Alford:2007xm}. 

In this subsection we will overview the basic results of \cite{Lawson:2015cla}
on neutrino and antineutrino fluxes from the sun. The main goal of ref.   \cite{Lawson:2015cla} 
was to argue that the Super-Kamiokande stringent constraint on anti-neutrino flux
 $\Phi_{\bar{\nu}_e} < (1.4-1.9){\rm cm^{-2} s^{-1}}$ at large energies  $E >19.3$ MeV 
   \cite{Lunardini:2008xd} (which played  the key role in analysis of \cite{Gorham:2015rfa}) does not apply to our case on neutrino and antineutrino production by AQNs
   because the typical energies  of neutrinos and antineutrinos will be much lower.

Indeed,  the  muons cannot be produced at all 
in the CFL phase for purely kinematical reasons. Therefore, the energetic antineutrinos  which are normally produced in 
the $\mu^{\pm}\rightarrow e^{\pm}\nu_e\nu_{\mu}$ decay channels are not   
produced  in the CS matter.  This is the crucial  point of the 
arguments presented in   \cite{Lawson:2015cla}.

In the simplest possible scenario (which was adopted in ref.  \cite{Lawson:2015cla}) the majority of neutrinos will be emitted by the lightest  
  NG bosons, in which 
 case the energy of the emitted neutrinos  does not normally  exceed 15 MeV for the CFL phase
because the lightest NG bosons do not normally  exceed mass in 30 MeV range as mentioned in previous subsection \ref{sec:NG}. This is very basic and very generic feature of the CS phase.
  We postpone the  important discussions on basic features of the neutrinos emitted by the quarks in CS phases to subsection \ref{nu-antinu}. Below we list   few   features   on the neutrino spectrum
if  it is saturated exclusively by NG bosons:
  
1.    A specific choice 
of $\mu$ and $\Delta$ determines the basic mass scales for the light NG modes, which eventually determine the $\nu$ spectrum;

2. It is normally assumed that a required value for $\mu$ for CFL to be realized is  $\mu \simeq 400$ MeV. The corresponding value for   $\Delta$ is estimated in this case as 
$\Delta \simeq 100$ MeV in the CFL phase, see review \cite{Alford:2007xm}.  Such  large value of  $\mu$  can be  indeed reached in the AQN's core as recent numerical studies suggest, see Fig. 2 in ref \cite{Ge:2019voa};

\exclude{3. The conventional Zweig suppression suggests that the heavier NG  $\pi^{\pm}$ bosons will be produced rather than
  lightest (for ani-nugget) $K^-$ boson. The corresponding suppression factor is estimated as $1/N_c^2\sim 0.1$, where $N_c=3$ is the number of colours in QCD;
}  
  3. The neutrino spectrum is qualitatively   different from that of the antineutrinos because the annihilation occurs not in vacuum but in dense CS state with  $\mu \simeq 400$ MeV.
\exclude{In particular, the relatively heavy $K^+$ will tend to the hadronic decays\footnote{This is because the leptonic decays are proportional to $m_e^2$
while hadronic decays  do not contain such suppression factor} producing 
a $\pi^+\pi^0$ rather than directly producing a neutrino. In contrast with light $K^-$ which may only decay as
$K^-\rightarrow e^-\bar{\nu}$.
Another difference leading to the distinct    spectra is that    $\pi^-$ in CFL phase can decay $\pi^-\rightarrow \bar{K}^0e^-\bar{\nu}$ to the light $\bar{K}^0$ in contrast with $\pi^+$ which must decay as $\pi^+\rightarrow e^+{\nu}$ because $K^0$ is heavy. These examples show   the main tendency here: the antineutrino spectrum is considerably less intense and characterized typically by lower energies  in comparison with the neutrino spectrum.
}
    
  4. Larger chemical potentials $\mu$ generally lead  to even lighter NG modes while the masses increase 
with the size of the gap $\Delta$ as eqs. (\ref{NG}) and (\ref{NG1}) state; 

\exclude{
6. Numerical estimates  presented in ref
    \cite{Lawson:2015cla} correspond to   very conservative 
  choice of parameters:   $\mu = 350$ MeV and 
$\Delta = 150$ MeV   when the lightest NG   are in 30 MeV range. 
For this choice of parameters the   neutrino's energies assume the highest possible value $\sim 15$ MeV. A choice with  $\mu = 450$ MeV (this value of $\mu$ is supported by the numerical computations  presented in  ref. \cite{Ge:2019voa}) leads to   smaller NG masses, and correspondingly   lower  neutrino's energies with $E_{\nu} \sim  10$ MeV rather than $E_{\nu}\sim 15$ MeV. This modest modification in scale does not change the basic features of the spectrum. 
}

We conclude this subsection with the following generic comment. The main goal of  \cite{Lawson:2015cla} was to argue (in simplified setting) that
the  neutrino's energies in the AQN framework are typically   in the 15 MeV range (in contrast with 20-50 MeV range as assumed in \cite{Gorham:2015rfa}) such that   the  stringent  constraint from  SuperK    \cite{Lunardini:2008xd}  does not apply.

 \subsection{Fermion excitations  and neutrino's production in CS phases.}\label{nu-antinu}
As we already mentioned  the neutrino emission from CS  phases quark matter have  been studied previously 
in a number of papers mostly in context of the physics of neutron stars, see  the review 
article \cite{Alford:2007xm}. We cannot literally apply the machinery developed in previous studies because
all excitations (such as NG bosons) in our case are produced not in a thermally equilibrium system when the density of 
of the excitations is unambiguously determined by the temperature. Instead, all excitations in our scenario  are produced as  a result of rare annihilation events.
These annihilation events excite the  NG bosons as well as fermion excitations, which may decay by emitting neutrinos. 
The coefficients $\kappa_{\nu}$ and $\kappa_{\bar{\nu}}$ entering (\ref{nu-anti-nu}) precisely correspond to this mechanism of  the neutrino production. 

In this subsection  we want to overview some features of the fermion excitations of the  CS phases 
which may be the dominant contributors  to  the neutrino fluxes. This is because the NG bosons which are produced as a result of annihilation events in CS phase cannot leave the system and consequently decay to emit neutrino, as they must stay inside of the nuggets. It should be contrasted  with hadronic phase when pions and Kaons (produced as a result e.g. $p\bar{p}$ annihilation) decay to muons and neutrinos in vacuum.   Therefore, the NG bosons in CS phases are likely to be absorbed by fermion excitations (if kinematically allowed), which consequently decay to neutrinos.

 We start our overview by mentioning some CS phases which support low energy fermion excitations.
 Detail discussions can be found in review article \cite{Alford:2007xm}.
 First of all, the so-called 2SC phase (when two out of three colours and flavours are paired and  condensed)  
 supports unpaired modes which could be light and couple to NG bosons. Another phase which also supports 
 the light fermion excitations is the so-called $CFL-K^0$ phase when $K^0$ energy vanishes according (\ref{NG1}).
 For antinuggets this  phase  corresponds to $\bar{K}^0$ condensation. In both cases (CFL and CFL-$K^0$) the gap  of the excitations decreases with increases of $p_F$ such that the gaps for $p$ and $n$ become lower. Indeed,  $\Delta_{p,n}=\Delta-\frac{m_s^2}{2p_F}$ such that $p, n$ fermion excitations could become completely ungapped. If this happens   these modes  may become the dominant producers of the neutrinos.  
 
 Indeed, in unpaired quark matter neutrino emissivity is dominated by the direct Urca processes  such as $d\rightarrow u+e^{-}+\bar{\nu}_e$. In case of antinuggets it should be replaced by anti-quarks with emission of neutrino $\nu_e$ and positron $e^+$ with the energy determined by the energy of the fermion  excitation, which itself assumes the energy of order of the lightest NG mode   according to  analysis  of the  previous section \ref{neutrino}. 
 
 If this process  indeed becomes the dominant mechanism of the neutrino emission from AQNs than one should expect that 
 \be
 \label{kappa}
 \kappa_{\nu}\gg \kappa_{\bar{\nu}}, ~~   E_{\nu}  \lesssim  15 ~{\rm MeV}, ~~~  \kappa_{\nu}\gtrsim 1,
  \ee
 which is assumed to be the case as Eq. (\ref{N_nu}) states.

\section{Local flashes}
\label{local flashes}
In this Appendix we generalize our  axion studies \cite{Liang:2019lya}  to include the neutrinos into consideration,  similar to what we have done in Section \ref{2}. The main 
topic for the present studies is the  enhancement effect and great amplification of the axion density which was coined as     a ``local flash'' in  \cite{Liang:2019lya}. It occurs  on rare occasions when an AQN hits (or exits) the Earth surface in vicinity of an axion search detector. In this Appendix we generalize the arguments of Section \ref{2} to estimate a similar local flash for neutrinos. 

\begin{figure}[h]
	\centering
	\captionsetup{justification=raggedright}
	\includegraphics[width=0.8\linewidth]{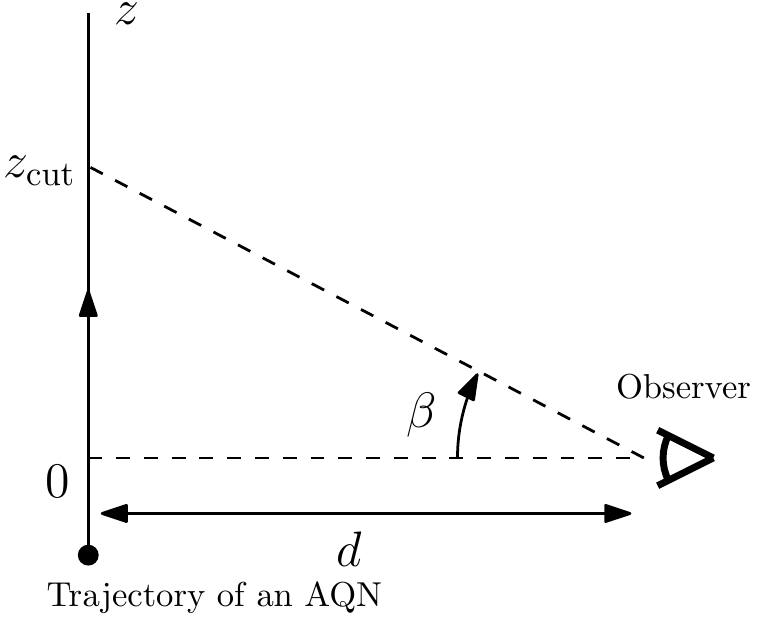}
	\caption{Local flash occurs   for  short period of time $\tau$   when an AQN moves at distance $d\ll R_{\oplus}$, adopted from \cite{Liang:2019lya}.}
	\label{fig:local flashes}
\end{figure}

We follow the same logic of ref \cite{Liang:2019lya} and 
 consider a case when an  AQN is moving in a distance $d$ close enough to the detector, as shown in Fig. \ref{fig:local flashes}.
 The total number of emitted neutrinos per unit area    within $z\in[-z_{\rm cut},z_{\rm cut}]$ as a result of passage of the AQN at distance $d$ from an observer is given by:
\begin{equation}
\label{nu}
\begin{aligned}
\Delta \frac{\rmd N_{\nu} }{\rmd A}
&=\frac{1}{4\pi}\int_{-z_{\rm cut}}^{z_{\rm cut}}
\frac{\kappa_{\nu}}{z^2+d^2}
\frac{\rmd m_{\rm AQN}(z)}{m_p}  \\
&\simeq\frac{\beta}{2\pi d}\frac{\kappa_{\nu}}{v_{\rm AQN}}\frac{\dot{m}_{\rm AQN}}{m_p}
\end{aligned}
\end{equation}
where $\beta$ is the angle related to $z_{\rm cut}$ as shown in Fig. \ref{fig:local flashes}, and we used formula (\ref{N_nu}) which defines the coefficient $\kappa_{\nu}$ as the number of neutrinos produced due to the   annihilation  of a single baryon charge. In obtaining (\ref{nu}) the integration $\rmd z$ is replaced by integration $\rmd m_{\rm AQN}$
 \begin{equation}
\label{nu1}
\rmd N_{\nu}\simeq \frac{\kappa_{\nu}}{m_p}\rmd m_{\rm AQN}
=\frac{\kappa_{\nu}}{m_p}\frac{\dot{m}_{\rm AQN}}{v_{\rm AQN}}\rmd z\ ,
\end{equation}
 Now we can estimate the flux of neutrinos due to the passage of this nearby AQN as follows 
\begin{equation}
\label{nu2}
\begin{aligned}
\Delta \left(\frac{\rmd N_{\nu} }{\rmd A \rmd t}\right)
\simeq \frac{1}{\tau}
\Delta \left(\frac{\rmd N_{\nu} }{\rmd A}\right)
=\frac{\kappa_{\nu}}{m_p}\frac{\beta}{4\pi d^2}\dot{m}_{\rm AQN}\ ,
\end{aligned}
\end{equation}
where we used expression (\ref{nu}) and approximated $\tau\simeq 2d/v_{\rm AQN}$ as a typical   travel time for an AQN inside  the interval $[-z_{\rm cut},z_{\rm cut}]$. 

 We want to  compare this ``local flash" (\ref{nu2}) with the average  flux \eqref{dN_nu} by introducing the amplification factor $A_{\nu}(d)$  defined as the ratio:
\begin{equation}
\label{nu3}
A_{\nu}(d)\equiv\frac{\Delta \left(\frac{\rmd N_{\nu} }{\rmd A \rmd t}\right)}{ \left(\frac{\rmd N_{\nu} }{\rmd A \rmd t}\right)}
\simeq\frac{\beta}{\langle \dot{N}\rangle\langle \Delta t\rangle}
\left(\frac{R_\oplus}{d}\right)^2\ ,
\end{equation}
where we   approximated $\dot{m}_{\rm AQN}\simeq\langle \Delta m_{\rm AQN}\rangle/\langle\Delta t\rangle$. The  physical meaning of 
$\langle\Delta t\rangle$ is  the time duration of the AQNs being averaged over all trajectories and   over the velocity distribution.  The result (\ref{nu3}) does not depend on neutrino's spectrum, nor intensity. It does not include even parameter $\kappa_{\nu}$, and in fact  identically coincides with expression obtained for the axion ``local flash"  derived previously  in \cite{Liang:2019lya}.
It is anticipated result as all these numerical factors cancel out in the ratio (\ref{nu3}) as relative amplification factor (\ref{nu3}) is entirely  determined by the dynamics of the AQNs, not the particles they emit as long as these particles are relativistic, which is the case for both species: the axions and neutrinos.  

As the final expression (\ref{nu3}) for the neutrino amplification factor coincides with the corresponding expression for the axion \cite{Liang:2019lya} the consequences in both cases are the same, and we simply list them: 

1. for  the typical ratio  $\langle \dot{N}\rangle\langle\Delta t\rangle\sim30$ where $\langle \dot{N}\rangle$ is given by eq. (\ref{eq:D Nflux 3 tot})   and  $\beta\sim1$ one can infer that an    amplification becomes significant if  $d\ll  0.1 R_\oplus$.

2. the time duration $\tau$ of a local flash as a function of amplification $A_{\nu}$:
\begin{equation}
\label{eq:tau}
\tau
\simeq\frac{2d}{v_{\rm AQN}} 
\simeq\left(\frac{\langle\Delta t\rangle}{\langle\dot{N}\rangle}\right)^{1/2}
\frac{1}{A^{\frac{1}{2}}_{\nu}}\ ,
\end{equation}
where in the last step, we approximate $v_{\rm AQN}\simeq2R_\oplus/\langle\Delta t\rangle$ and assume $\beta\sim1$ for simplicity. We summarize a few choices of time duration $\tau$ as a function of amplification factor $A_{\nu}$ in  Table \ref{tab:local flashes}. 
  
3.   The probability to observe an AQN for $z\leq d$ behaves as a simple area law:
\begin{equation}
\label{eq:Prob(z_min=d)}
{\rm Prob}(z \leq d)
\simeq\left(\frac{d}{R_\oplus}\right)^2
\simeq\frac{1}{\langle \dot{N}\rangle\langle \Delta t\rangle}\cdot\frac{1}{A_{\nu}}, \ .
\end{equation}
where we use Eq. (\ref{nu3}) to express $d$ in terms of $A_{\nu}$. 

4. the event rate can be expressed in terms of amplification parameter  $A_{\nu}$,
\be
\label{eq:Event rate}
{\rm Event~rate}
=\frac{\la\dot{N}\ra\cdot {\rm Prob}(z \leq d) \cdot \tau}{\la\Delta t\ra}
\simeq\frac{A_{\nu}^{-3/2}}{\sqrt{\langle\dot{N}\rangle\langle\Delta t\rangle^3}},~~~~~~
 \ee
where averages  $\langle\dot{N}\rangle$ and $\la\Delta t\ra$ have been numerically computed for different size distribution models in \cite{Liang:2019lya}.


\bibliography{DAMA-LIBRA}

\end{document}